# Uncertainty Quantification of Water Distribution System Measurement Data based on a Least Squares Loop Flows State Estimator

Corneliu T.C. Arsene
corneliuarsene@gmail.com
School of Science and Technology
Nottingham Trent University
Nottingham, NG11 8NS, United Kingdom

*Abstract*—This paper presents a novel algorithm for uncertainty quantification of water distribution system measurement data including nodal demands/consumptions as well as real pressure and flow measurements. This procedure, referred to as a Confidence Limit Analysis (CLA), is concerned with a deployment of a Least Squares (LS) state estimator based on the loop corrective flows and the variation of nodal demands as independent variables. The confidence limits obtained for the nodal pressures and the inflows/outflows of a water network are determined with the novel algorithm called Error Maximization (EM) method and are evaluated with respect to two other more established CLA algorithms based on an Experimental Sensitivity Matrix (ESM) and on the sensitivity matrix method obtained with the LS nodal heads equations state estimator. The estimated confidence limits obtained for two real water networks show that the proposed EM algorithm is comparable to the other two CLA benchmark algorithms but due to its computational efficiency it is more suitable for online decision support applications in water distribution systems. Both ESM and EM methods work for any operating point whether arbitrarily or randomly chosen for any water network although EM method has the advantage of being computationally superior and working with any sets of measurements.

*Keywords-uncertainty quantification, confidence limit analysis, water distribution systems, measurement data, loop corrective flows algorithms, modelling and simulation, state estimation*

## I. Introduction

The existence of water distribution systems has been extremely important in the human history before even the Biblical times and more than 4000 years ago. Modern water distribution systems are large, complex and highly non-linear in their behaviour. They include many complex elements and structures like pipes, pumps, valves, reservoirs which are all interconnected into distribution networks serving consumers of different type ranging from large factories to individuals.

In order to provide water to consumers without any disruption in service, the state of the water distribution system has to be monitored. This can be achieved by using a mathematical model of the system, the so called state estimator, which provides a means of combining diverse measurements (e.g. nodal consumptions/demands, pressures, pipe flows) and information about the distribution network topology and features [1-7].

Although the mathematical model may be accurate, the state estimates are based on input data that may contain a significant amount of uncertainty [1][8-11]. The uncertainty in input data associated with the real measurements of flows and pressures, as well as the pseudo-measurements (e.g. an estimation of the water consumption at nodes) [12-14], is discussed here in the context of a Least Squares (LS) loop flows state estimation technique [1][8][15-16]. There are other types of uncertainties that can be found in water distribution systems such as the uncertainties in the pipe roughness coefficients [17-20] (i.e. water network parameters) which however will not be taken into account directly in our analysis.

This paper addresses the measurement data uncertainty which has an impact on the accuracy with which the state estimates are calculated [8-10]. It is important, that the system operators are given not only the values of flows and pressures in the network at any instant of time but also that they have some indications of how reliable these values are. In the context of water distribution systems, the procedure for the quantification of the inaccuracy of the state estimates caused by the input measurement data uncertainty was developed in the late 1980s and referred to as a Confidence Limit Analysis (CLA) [9][21][22]. Rather than a single deterministic state estimate, the CLA enables the calculation of a set of all feasible situations of a water network corresponding to a given level of measurement uncertainty [9][21]. The set is presented in the form of upper and lower bounds for individual water network variables and hence provide limits on the potential error of each water network variable. A decision support system built using the concept of CLA has been further developed in [10][21]. It performed like a fault detection and identification system being able to distinguish between different types of errors that are occurring in water networks. It should also be mentioned that more general mathematical methods for quantification of uncertainties in the input measurement data or in the parameters of a system have also been developed [23][24] and have a widespread use in various engineering areas.

Although a great amount of work has been done, and significant results have been delivered in the area of uncertainty analysis of water networks, most of these algorithms were obtained with the nodal heads network equations [4][9][21][25]. This raises the question of the potential benefits of using the loop equations for CLA. Employing the loop corrective flows variables for the numerical simulations or state estimation procedures has received an increased attention [2][3][15][16]. As it has been shown in the hydraulics literature, satisfactory convergence and good numerical stability have been achieved for the loop corrective flows based simulations [1][2][15][16] in contrast to some potential instabilities encountered when using the nodal heads based equations. In spite of the work in water networks state estimation and simulation, the results in the area of CLA when using the loop flows algorithms are limited [3][8].

This paper extends the author' previous work [2][8][15][16][26][27] on a Least Squares (LS) loop flows state estimator for water networks to include the CLA and arrive at state estimates with the respective confidence limits. In [28] there were presented briefly two CLA algorithms based on a LS loop flows state estimator: a) the Experimental Sensitivity Matrix (ESM) method, and b) a novel Error Maximization (EM) method. The confidence intervals estimated by these two CLA algorithms were compared with the confidence intervals estimated with a CLA algorithm which was using the nodal heads equations for the implementation of the LS state estimation procedure. The performances of the ESM and the novel EM algorithms were assessed with respect to computational convergence time required by each of the evaluated algorithms. In this paper, the descriptions of the two novel CLA algorithms are enlarged and the algorithms are well placed in the context of the other available works in the CLA domain. The respective CLA algorithms are also preceded by the full presentation of the LS loop flows state estimator for a clear understanding of the novel CLA methods. Finally, one of key points of this paper is the application of the EM method on a new and important large water network, which will prove the effectiveness of this method.

The paper is organized as follows. Section II presents the numerical algorithms used for the simulation and state estimation of a water network based on the loop corrective flows variables. Section III presents the background of the uncertainty quantification of the water networks measurement data. This is followed by Section IV describing the calculation of the ESM. Section V presents the CLA based on the ESM method which is used as a benchmark for interpreting the results obtained in the following section with the proposed, novel EM technique applied to two different real water networks. Finally the summary and conclusions are presented in the last section.

## II. Numerical Algorithms

Two main numerical algorithms based on the loop corrective flows and the variation of nodal demands variables are used in this paper: a) a water network simulator; and b) a state estimator.

*A. Water Network Simulator Algorithm*

Modeling and simulation of water distribution system consists of two main ingredients: the set of independent equations that describe the water network and the numerical optimization method used to calculate the nodal heads and the pipe flows. There are three main ways of constructing the network equations: the flows [29], the nodal heads [30][31], and the loop corrective flows [32-36]. More details about each of these simulation methods can be also found in [37-41]. Once the network equation has been established, it is solved iteratively with the Newton-Raphson numerical method. However, several numerical methods can be used to solve the mathematical model of a water network and they can be classified in the following major groups: a) the numerical minimization methods [42][43]; b) the Hardy-Cross method [44][45]; c) the Newton-Raphson method [46-48]; and d) the Linear Theory method [49]. There have been also various algorithms developed for solving water networks containing hydraulic controlling elements such as check-valves, pressure reducing valves or pumps [50-65].

There are various types of simulations that a water network model may perform, depending on what the modeler is trying to observe or predict. The two most basic types are: *steady-state simulation* [66-68] which computes the state of the water distribution system (flows, pressures, valve position, etc.) assuming that the water demands and boundary conditions do not change with respect to time or *extended period simulation* [69-71] which determines the dynamic behaviour of a system over a period of time by computing the state of the water distribution system as a series of steady-state simulations in which water demands and boundary conditions do change with respect to time. Other types of simulations are *water quality simulations* [72-76] used to ascertain chemical or biological constituent levels within a system or to determine the age or source of water, automated fire flows analyses [77][78] that establish the suitability of a system for fire protection needs or cost analyses that are used for looking at the monetary impact of operations and improvements.

In Fig. 1 a simple example of a water network is shown where the edges are the pipes that distribute the water to the consumers represented by the nodes (e.g. 1, 2, 3, etc.). A simulator algorithm is defined as a solution of the water network equations for a given set of nodal demands (i.e. steady state simulation). The variables $\Delta Q_{l1}$ , $\Delta Q_{l2}$ and $\Delta Q_{l3}$ (shown in Fig. 1) are some loop corrective flows which will be described later. Nodes 1 and 4 (represented as squares) are the nodes with fixed pressure which are called the fixed-head nodes.

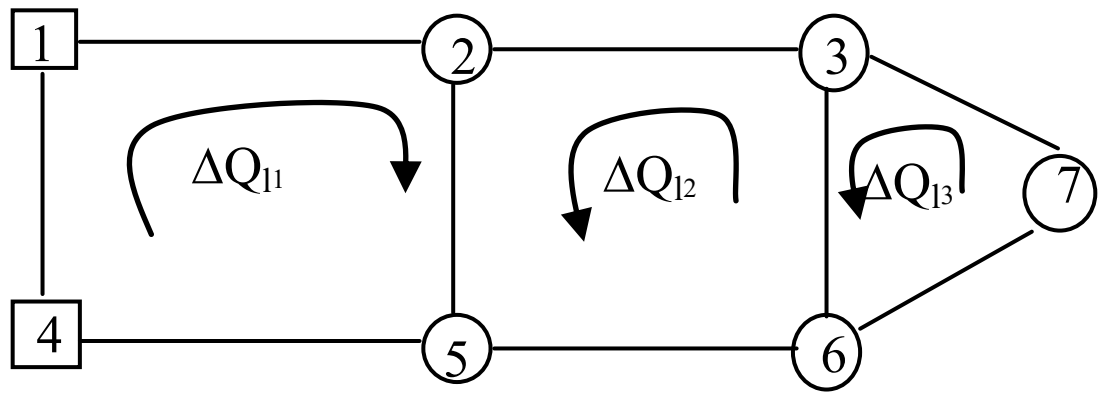

Figure 1. Example of a simple water network.

The simulator algorithm used here is based on the loop corrective flows algorithm defined for a water distribution system with n-nodes, l-loops, and p-pipes. The continuity equations which states that the flow entering a node must equal the nodal consumption plus the flow exiting the respective node, must be satisfied. Therefore, an initial *i*-th pipe flows solution $\boldsymbol{Q}_i$ (*p* x *1*) that satisfies the continuity equation is calculated as:

$$\boldsymbol{A}_{np}\boldsymbol{Q}_i=\boldsymbol{d} \qquad (1)$$

where $\boldsymbol{d}$ are the nodal demands (*n* x 1) and $\boldsymbol{A}_{np}$ is the topological incidence matrix (*n* x *p*).

The topological incidence matrix $\boldsymbol{A}_{np}$ has a row for every node and a column for every pipe of the water network. The entries for each row, which can take a value of +1 or –1, indicate that the flow in a pipe enters or leaves the node [1][8]. The mass balance equation is solved with the Newton-Raphson method and states that the sum of the pipe head losses around each loop must be equal to 0:

$$\boldsymbol{M}_{lp}\,\boldsymbol{h}=\boldsymbol{0} \qquad (2)$$

where $\boldsymbol{h}$ represents the pipes head losses (*p* x 1) calculated by the Hazen-Williams equation [2] and $\boldsymbol{M}_{lp}$ is the loop incidence matrix (*l* x *p*).

The entries of the loop incidence matrix can take the value of: +1 when the flow in a pipe has a clockwise direction, -1 for the anti-clockwise direction of the flow, and 0 when a pipe does not belong to a loop [1][8].

The loop corrective flows $\mathbf{\Delta Q}_l$ ($l$ x 1) at the step k+1 of the Newton-Raphson iteration method which solves Eq. (2) are:

$$\Delta Q_{l_{k+1}} = \Delta Q_{l_k} - \left[ \frac{\partial \Delta H}{\partial \Delta Q_{l_k}} \right]^{-1} \Delta H \qquad (3)$$

where $\Delta \boldsymbol{H}$ are the residual loop head losses ($l$ x 1) (i.e. $\Delta \boldsymbol{H} = M_{lp} \boldsymbol{h}$), $k$ is the step of the Newton-Raphson optimization method.

The Jacobian matrix $\frac{\partial \Delta H}{\partial \Delta Q_{l_k}}$ in (3) ($l$ x $l$) can be expressed as:

$$\boldsymbol{J} = \boldsymbol{M}_{lp} \boldsymbol{A} \, \boldsymbol{M}_{pl} \qquad (4)$$

where $\boldsymbol{M}_{pl}$ is the transpose of loop incidence matrix $\boldsymbol{M}_{lp}$ ($l$ x $p$) and $\boldsymbol{A}$ is a diagonal matrix with a special property ($p$ x $p$):

$$A = \begin{pmatrix} s k_1 |Q_1|^{s-1} & 0... & 0 \\ 0... & s\, k_2 |Q_2|^{s-1} & 0 \\ 0 & 0... & s\, k_p |Q_p|^{s-1} \end{pmatrix} \qquad (5)$$

where $k_{1,2,..p}$ is the pipe head loss coefficient and $s$ is the exponent in the Hazen-Williams equation [2].

The final pipe flow solution $\tilde{Q}$ for each pipe is:

$$\tilde{Q} = Q_i + M_{lp}{}^T \Delta Q_l \qquad (6)$$

where $\tilde{Q}$ (p x 1) are the final pipe flows calculated at the end of the Newton-Raphson method [1][2][8].

The loop simulator requires the computation of the loop incidence matrix $\boldsymbol{M}_{lp}$ and the initial pipe flows $\boldsymbol{Q}_i$. This problem is in general based on the decomposition of the water network into a spanning tree. A spanning tree contains all the vertices and the edges of a connected and undirected graph except for the edges which form the cycles (i.e. loops) of the graph [1][8]. A spanning tree for the simple water network from Fig. 1 is shown in Fig. 2.

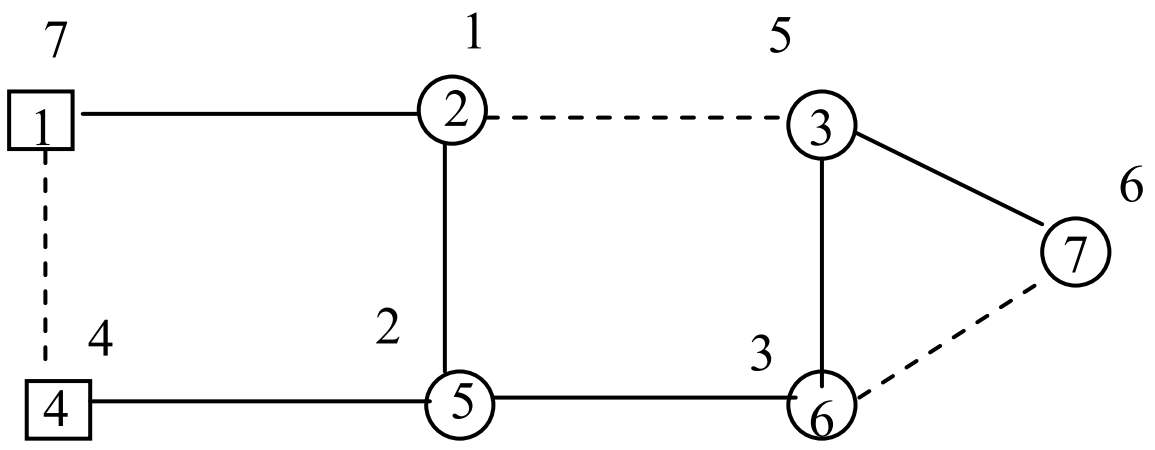

Figure 2. Spanning tree associated to water network where the dashed line close the loops and the renumbering of the nodes and pipes produces an upper triangular tree incidence matrix *T* which describes the incidence of nodes and pipe from the spanning tree.

The construction of a spanning tree starts usually from a node which becomes the main source node. Different search strategies can be employed in order to search the water network. For example, the Depth First (DF) search from the graph theory can be one of the choices for finding the loops in a water network. It has the property that always a pipe that does not belong to the spanning tree (so called a chord pipe) connects a node with one of its predecessors in the tree. In Fig. 2 the pipes which close the loops are shown with dashed lines while the main source node is the node 1 from Fig. 1 which now becomes node 7 in Fig. 2.

Based on the spanning tree, the topological incidence matrix $\boldsymbol{A}_{np}$ can be split into an upper triangular tree incidence matrix *T* which defines the incidence of the tree pipes, which are the pipes situated in the spanning tree, and a co-tree incidence matrix $\boldsymbol{C}$ which contain the pipes that are not in the spanning tree and are called co-tree pipes or chords and form the loops (i.e. $\boldsymbol{A}_{np} = [\boldsymbol{T}\ \boldsymbol{C}]$). A renumbering of nodes and pipes might be necessary in order to obtain the upper triangular tree incidence matrix (Fig. 2). In this case the loop corrective flows are the co-tree pipe flows. Also between each of the fixed-head node and the main source node is added a loop which is called a pseudo-loop and a loop corrective flow (i.e. co-tree flow) is considered for the respective pseudo-loop.

*B. LS loop flows state estimator*

The application of the LS or Weighted LS (WLS) state estimators for on-line monitoring of water networks has been extensively studied by a number of researchers. In most of these studies [21][25][79] the nodal heads have been used as independent variables. In contrast to these studies, in [3] the independent variables were constructed using chord flows (i.e. a particular type of loop equations) and a constrained optimization problem was solved using a Successive Quadratic Programming (SQP) technique which implied the calculation of the Jacobian and the Hessian matrixes of the objective function and the Jacobian of the reduced set of constraints. In [80] the chord flows and the pipe roughness coefficients were used as independent variables and again the SQP technique was used. In [81] the chord flows and pipe diameters were estimated using the SQP method. The use of more than one set of independent variables in building the mathematical model of a distribution network was also advocated in [82] for gas distribution networks where the continuity equations and the loop equations were solved together. A similar approach for water networks was discussed in [83] where again the continuity equations and the loop equations (i.e. expressed as a function of the two nodal heads ends of the chord pipe) were solved simultaneously.

In this paper, we propose an additional set of independent variables for the state estimation procedure which represent the variation of nodal demands $\Delta \boldsymbol{d}$ ($n$ x $1$). In this case the pipe flows are written with respect to the loop corrective flows $\Delta \boldsymbol{Q}_l$ and the variation of nodal demands:

$$\tilde{Q} = \boldsymbol{Q}_i - \boldsymbol{A}^* \Delta \boldsymbol{d} + \boldsymbol{M}_{pl} \Delta \boldsymbol{Q} \qquad (7)$$

where $\tilde{Q}$ ($p$ x 1) are the flows in a tree and co-tree pipes (p x 1), $Q_i = \begin{bmatrix} Q_{T_i} \\ 0_l \end{bmatrix}$ ($p$ x 1) are the initial flows determined from the spanning tree for the tree pipes while the initial flows in the co-tree pipes are zero (i.e. $0_l$ zero vector of size (1 x $l$) ), and matrix $\boldsymbol{A}^*$ ($p$ x $n$) is the matrix with the property $A^* = \begin{bmatrix} T^{-1} \\ 0_{\text{ln}} \end{bmatrix}$ where $\boldsymbol{0}_{ln}$ is the zero matrix of size ( $l$ x $n$ ).

There are two sets of equations which are used to describe the hydraulics of the water network. The first set of equations states that the loop head losses around the loops equal to zero:

$$\Delta H(\Delta Q_l, \Delta d) = 0 \quad (8)$$

where the loop head losses residuals $\boldsymbol{\Delta H}$ are function of the loop corrective flows $\boldsymbol{\Delta Q}_l$ and the variation of nodal demands $\boldsymbol{\Delta d}$.

The second set of equations states that the total amount of inflow/outflow to/from the water network carried out through the fixed-head nodes is equal to the variation of nodal demands. This results from the mass balance law at each node and can be written as:

$$\boldsymbol{\Delta d} = \boldsymbol{B}_{nl}\,\boldsymbol{\Delta Q}_l \quad (9)$$

The matrix $\boldsymbol{B}_{nl}$ ($n$ x $l$) has non-zero elements equal to 1 which corresponds to the main root node and -1 for each of the fixed-head nodes.

The equations (8) and (9) represent the hydraulic function that describes the water network. It can be written as the following system of equations:

$$\begin{cases} \Delta H(\Delta Q_l, \Delta d) = 0 \\ B_{nl}\Delta Q_l - \Delta d = 0 \end{cases} \quad (10)$$

The system of equations (10) can be solved using the Newton-Raphson iterative method and minimizing the LS criterion which assures the existence of a single and global optimum point. The Jacobian matrix of the system of equations is

$$J = \begin{bmatrix} \dfrac{\partial(\Delta H)}{\partial(\Delta d)} & \dfrac{\partial(\Delta H)}{\partial(\Delta Q_l)} \\ -\dfrac{\partial(\Delta d)}{\partial(\Delta d)} & B_{nl}\dfrac{\partial(\Delta Q_l)}{\partial(\Delta Q_l)} \end{bmatrix} \quad (11)$$

There are two slightly different versions of the same state estimator and one can use any of the two versions: a) a state estimator which uses the Eq. (11) and requires the calculation of the few elements of matrix $\mathbf{B}_{nl}$ ; and b) a state estimator which uses Eq. (12) and where matrix $\mathbf{B}_{nl}$ is zero. In case of Eq. (12)

the inflows/outflows of the water network are calculated at the end of the Newton-Raphson method by subtracting the variation of the nodal demands at the fixed-head nodes from the loop corrective flows corresponding to the pseudo-loops (i.e. loops added between the main root node and each of the fixed-head nodes). In this second case, the Jacobian matrix of the LS loop flows state estimator can be written as:

$$J = \begin{bmatrix} \frac{\partial(\Delta H)}{\partial(\Delta d)} & \frac{\partial(\Delta H)}{\partial(\Delta Q_l)} \\ -I_{nn} & 0_{nl} \end{bmatrix} \qquad (12)$$

where $\mathbf{I}_{nn}$ is the ($n$ x $n$) identity matrix.

The Jacobian matrix $J$ resembles the one presented in [1][8][15][16][26][27]. It is paramount to underline the followings which were not discussed previously in connection to this LS loop flows state estimator which uses two sets of independent variables: the loop corrective flows and the variation of nodal demands $\boldsymbol{\Delta d}$. The new nodal demands $\boldsymbol{d_f}$ are calculated based on the values of $\boldsymbol{\Delta d}$ obtained with the LS loop flows state estimator:

$$\boldsymbol{d_f = d - \Delta d} \qquad (13)$$

where $\boldsymbol{d}$ are the nodal demands before the state estimation, $\Delta d$ is the variation of nodal demands obtained with the LS loop flows state estimator and $\boldsymbol{d_f}$ are the new nodal demands.

Once the new nodal demands $\boldsymbol{d_f}$ are obtained, there could be necessary a second computational step for the situations where it can happen that the LS loop flows state estimator to calculate a variation of nodal demands which eventually to result in $\boldsymbol{d_f = d - \Delta d < 0}$ . However, this is not a problem as the negative part in the demand values in the vector $d_f$ can be regarded as residuals for the respective nodal demands and the respective nodal demands will be taken as being zero in a second computational step which will consist of a loop simulation in which the nodal demands $\boldsymbol{d_f}$ equal or higher than zero will be used. Therefore during the second computational step, the nodal demands are considered to be known and obtained from the previous step, that is the state estimation, whereas once again the negative values in the vector $d_f$ will be taken as being zero. Finally, another procedure for dealing with the situation in which $d_f = d - \Delta d < 0$ is presented in [26] where the LS loop flows state estimator was iteratively constrained so that the nodal demands $\boldsymbol{d_f}$ not to become smaller than zero during the Newton-Raphson iteration method but however this second procedure can result sometimes in lack of convergence so this is why the one presented herein is preferable.

This LS loop flows state estimator was extensively tested while obtaining the correct physical water network solutions for medium size water networks and for various variations in a set of measurements including water network consumptions, fixed-head nodes, variable nodal heads and pipe flows.

## III. Uncertainty Analysis of Water Networks Measurement data

In numerous water distribution systems state estimation techniques, it has been demonstrated that for a specific combination of input data and estimation method there is one optimal solving [1][2][4-6].

However, because of the inexactness in the input data, there are many possible and different mixtures of such input data and therefore there are many correct but yet dissimilar calculated state estimate vectors.

As a consequence, the uncertainty studies turn out to be unavoidable constituents of the water networks modeling and simulation because it is paramount, for the avoidance of any risks from the system operational control point of interest, to have knowledge about how the input data inexactness can influence the state estimates.

A very large amount of work on the implications of measurements (nodal heads, pipe flows) and pseudo-measurements (nodal consumptions) inaccuracies in water network has been realized [4][9][10][21][84]. The developed methods and the resulted studies were built on the axiom of unknown-but-bounded errors for the entire group of measurements:

$$\boldsymbol{z}=g(\boldsymbol{y})+\boldsymbol{r}\ ,\ |r_i| \leq |e_i|\ ,\ i=1,\ldots,m \tag{14}$$

where $\boldsymbol{e}$ is a vector describing the maximum expected measurement errors, $\boldsymbol{z}$ is the measurement vector, $g$ is the hydraulics water distribution system function, $\boldsymbol{y}$ is a set of LS state estimated variables (i.e.

independent variables) used to build the mathematical model from Eq. (14), $\boldsymbol{r}$ is the vector of residuals which can not be justified by the state estimator and the measurement data.

For a state estimator based on the nodal heads equations as in [9][21], the measurement vector $\boldsymbol{z}$ comprises the nodal demands, which are also called pseudo-measurements, and the real measurements which can be any mixture of real-life pressure and flow measurements. The information about the statistical properties of errors may not be necessary and the only restriction imposed was that of the errors falling within a range bounded by $\boldsymbol{e}$ [4][9][21]. Several CLA techniques have been put forth in the context of simulation and state estimation of water distribution systems based on nodal heads equations, but the most successful ones in terms of computational complexity have been based on the linearized model of the water network [9][10][21]. The linearized model of the water network was employed to deliver a sensitivity matrix $\boldsymbol{S}$. The sensitivity matrix was the pseudo-inverse of the Jacobian matrix calculated for the state estimates by using the LS nodal heads state estimator [9][10][21].

It is paramount to underline that in the context of the LS nodal heads state estimator, the LS state/independent variables are the nodal heads/pressures variables which are the same variables as the actual water network state variables of interest.

An LS nodal heads state estimated vector was determined on the grounds that the true measurement vector $\boldsymbol{z}^t$ (nodal demands/pseudo-measurements and the few other real pressure and flow measurements) is free of errors and the possible maximum error of the measurement set $\Delta\boldsymbol{z}$ was considered together with the sensitivity matrix $\boldsymbol{S}$ in order to estimate the confidence limits in the LS nodal heads state estimated vector. This method was possible to be implemented because of the use of the nodal heads equations in the LS nodal heads state estimator. Because of this, the $(i,j)$-th constituent $s_{ij}$ of the pseudo-inverse of the Jacobian matrix (i.e. matrix $\boldsymbol{S}$) relates the sensitivity of the $i$-th constituent, $y_i$, of the true LS nodal heads state vector , $y^t$, to the $j$-th constituent, $z_j$, of the measurement vector.

In order to stress the last idea, in the situation of the LS nodal heads state estimator, the LS nodal heads state estimates are the current water distribution system state variables which are the nodal pressures and the inflow/outflows from the water distribution system as well as some real nodal heads and pipe flow measurements placed for instance by the human operator.

Determining the confidence limits for the LS nodal heads state estimates $y_i$ , which were the nodal head/pressures or the few real measurements comprising of pipe flows or nodal head measurements, was calculated in this sensitivity matrix technique as:

$$y_{cl_i} = max(\, \boldsymbol{s}_p \, \Delta \boldsymbol{z} \,) \qquad (15)$$

where $\boldsymbol{s}_p$ is the $p$-th row of the sensitivity matrix $\boldsymbol{S}$, $y_{cli}$ is the confidence limit for the $i$-th LS nodal heads state estimates, $\Delta \boldsymbol{z}$ represents the maximum perturbations/errors in the vector of measurements.

The main principle of the CLA is to assume the worst situation for the inaccuracies in the vector of measurements (i.e. the maximum variation of nodal demands and errors for real pressure or flow meters).

Therefore the sensitivity matrix technique was very good and appropriate since in the LS nodal heads state estimator any mixture of real measurements could be employed together with the nodal demands with the scope of obtaining the confidence limits on the LS nodal heads state estimates/variables (i.e. nodal heads and pipe flows).

However, other methods have been investigated in the literature which are not all taking into consideration the nodal heads equations in constructing the mathematical model for the calculation of the confidence limits for the water distribution system state variables (i.e. nodal pressures and pipe flows).

In [85] a fuzzy method which conveyed fuzzy equations into a nonlinear optimization statement, was investigated for the calculation of confidence intervals and it was built on top of both the continuity equations and the loop head losses equations which were used to implement the water network function, while the Lagrangian technique was employed to determine the minimum of the objective function.

The same as in [85], fuzzy state estimation for water distribution systems was developed in [86][87].

Fuzzy reasoning, for instance for characterizing the nodal head values, has been also employed in [88-90], while in [91] fuzzy reasoning was mixed with Genetic Algorithm (GA) for multiobjective optimization of water distribution systems.

In [12][92] the implications of the nodal consumptions inaccuracies on the nodal pressures was considered by employing an approximation of the standard deviation of the nodal pressures values and which was represented as a relationship of the nodal consumptions.

Nodal consumptions inaccuracies have been also investigated in [93][94] while in [95] both the inaccuracies in measurement data and the parameters have been investigated based on the modeling with the nodal pressures equations and linear fractional transformations.

However, inaccuracies in the nodal consumption measurements or the pipe roughness variables, which are considered to be the water network parameters not similar to the water distribution system measurements, can be also considered when building reliability measures for water distribution systems with the scope of assuring for example some minimal pressure requirements [96].

It should be stressed that the usual techniques for taking into consideration the inaccuracies in the water distribution system measurement data (i.e. nodal consumptions, nodal heads or flow measurements) or water distribution system parameters (i.e. pipe roughness parameters) have been the sampling methods such as Monte Carlo method [97-101], Latin Hypercube sampling method [102] or Metropolis method [19].

Kalman filtering method was also employed to calculate the pipe roughness parameters and their variability [103] while specific reliability methods (e.g. first-order second moment technique) were employed to take into consideration the inaccuracies in the water distribution system variables [104-106].

Within the framework of the Gradient Algorithm [34] for simulating water distribution systems, the inaccuracies in the pipe roughness variables were taken into consideration and confidence limits were calculated for the nodal consumptions with a technique similar to the sensitivity matrix technique and which was named the grey numbers technique [107].

Uncertainty studies of measurement data and/or input variables has been also realized in other, connected domains such as electrical power systems [108][109] or civil structures [110].

The sensitivity matrix based CLA technique presented earlier, is not easy to use straightforward within the loop equations formalism. The equation which determines the loop corrective flows in a simulator algorithm [1][8][15][16]:

$$\Delta Q_l = -\left[\frac{\partial \Delta H}{\partial \Delta Q_l}\right]^{-1} \Delta H \qquad (16)$$

where element $s_{ij}$ of the inverse of the Jacobian matrix $\left[\frac{\partial \Delta H}{\partial \Delta Q_l}\right]^{-1}$ relates the sensitivity of the *i*-th element, $\Delta Q_{l_i}$, of the vector of loop corrective flows, $\Delta Q_l$, to the *j*-th element, $\Delta H_j$, of the vector of loop head losses residuals.

The intrinsic characteristics of the inverse of the Jacobian matrix from (Eq.16) which is making the connection between the loop head losses and the loop corrective flows is not the same as to the relationship from the nodal heads equations simulator/estimator method.

For a random variability in the measurement vector, it will have to be used directly the numerical values of the loop head losses residuals (i.e. vector $\Delta H$) and which to be used in order to determine the variability of the loop corrective flows. However, these loop head losses are determined with non-linear equations which describe the summation of the head losses in pipes belonging to a loop. In consequence, for a variability in the vector of measurements (i.e. water nodal demands, real-life nodal and pipe flow measurements) it is not possible to employ immediately equation (16) with the scope of calculating the confidence limits for the water distribution system state variables of interest which would be for instance the nodal pressures.

However, in the context of the loop equations, in [83][111] it was built a state estimation method and uncertainty study of the input measurement data, which were resulting in a sensitivity matrix method which was using the loop equations and continuity equations (i.e. mass balance equations).

In the following chapter it is presented an experimental sensitivity matrix (ESM) which is using the LS loop flows state estimator. As it will be shown in the following section, the ESM has the same characteristics as the inverse of the Jacobian matrix from the LS nodal heads equations simulator algorithm.

## IV. Experimental Sensitivity Matrix

Usually, the state estimators obtain a vector of water network state variables corresponding to a single measurement set [1][2][14]. Therefore the deterministic state estimators do not show any insight with regard to the ways in which the water network variables might be influenced by the inexactness in the input data. On other hand, if a deterministic state estimator is employed a number of times for each measurement which is changed function of its maximum associated error interval, afterwards an Experimental Sensitivity Matrix (ESM) $S^e$ can be calculated as:

$$S^e = \frac{\Delta x_i}{\Delta z_j} \quad i=1,\dots n; j=1,\dots m \qquad (17)$$

where $i$ is the index for the water distribution system state vector $\mathbf{x}^t$ (nodal heads and in/out flows), $j$ is the index for the measurement vector $z^t$ (nodal water consumptions), $\Delta x_i$ is the variation in the $i$-th element, $x_i$, of the true water distribution system state vector, $\mathbf{x}^t$, $\Delta z_j$ is the variation in the $j$-th element, $z_j$, of the true measurement vector $z^t$, $S^e$ is the Experimental Sensitivity Matrix (ESM).

The measurement vector $z^t$ includes the estimates for the nodal consumptions and the fixed-head nodal values.

The LS loop flows state estimator is the deterministic state estimator employed to calculate the ESM.

The matrix $S^e$ is actually the ESM because it has the properties of the inverse of the Jacobian matrix from the nodal heads simulator algorithm.

The ESM describes the variation, $\Delta x$, of the $i$-th constituent, $x_i$, of the true water network state vector, $x^t$, because of an error, $\Delta z$, in the $j$-th constituent, $z_j$, of the true measurement vector $z^t$.

The ESM is determined by carrying out several simulations which number of simulations is the same as the number of water network nodal consumptions and fixed-head nodes.

The true state of the water distribution system might not be available but it is possible to use an appropriate water network state vector $\hat{x}$ which it is known and it can be employed to calculate the ESM and the confidence limits.

The ESM technique is carried out with regard to the water distribution system displayed at Fig. 3. The data characterizing the water distribution system is summarized in the Appendix 1.

For each pseudo-measurement (i.e. nodal water consumption), a closed set is considered $[z_l \; z_u]$ in agreement with the relative variability of $z^t$. The error bounds $z_l$ and $z_u$ are situated at the same distance with regard to $z^t$ (i.e. $z^t - z_l = z_u - z^t$ ).

The variations of the water network nodal demands used is ±20% and the accuracy of the fixed-head nodes is ±0.01 (m).

In real-life water networks, the true measurement vector $z^t$ is very seldom similar to the observed measurement vector $z^o$.

This difference is because of the existence of errors in meters when dealing with the real measurements or because of the difficulties in estimating the consumptions when dealing with the nodal consumptions. Therefore the observed measurement values from the Table 1 and Table 2, are not similar to the true measurement values that could be met in the true operating state and which are shown on the 2-nd column of Table 1 and the 2-nd and 5-th column of Table 2.

The observed measurement values, $z^o$, are picked up randomly from within the interval determined by the upper $z_u$ and the lower limits $z_l$:

$$z_l = z^t - \Delta z_l \qquad (18)$$

where $\Delta z_l$ is the maximum variability of the lower level of the true measurement vector, $z_l$ is the lower level of the true measurement vector.

$$z_u = z^t + \Delta z_u \qquad (19)$$

where $\Delta z_u$ is the maximum variability of the upper level of the true measurement vector, $z_u$ is the upper level of the true measurement vector.

An error interval has just been calculated for the true measurement vector, $z^t - \Delta z_l \le z^t \le z^t + \Delta z_u$ (i.e. usually $\Delta z_l = \Delta z_u$), which can be met in the real-life events where measurement values are not précised but are situated in a domain defined by the exactness of the real meters and the exactness of the water network nodal consumptions. The water distribution system state vector $\hat{x}$ displayed in columns 3 and 6 of Table 3 includes the water distribution system state variables (i.e. nodal heads and in/out flows) determined for the observed measurement vector with the LS loop flows state estimator.

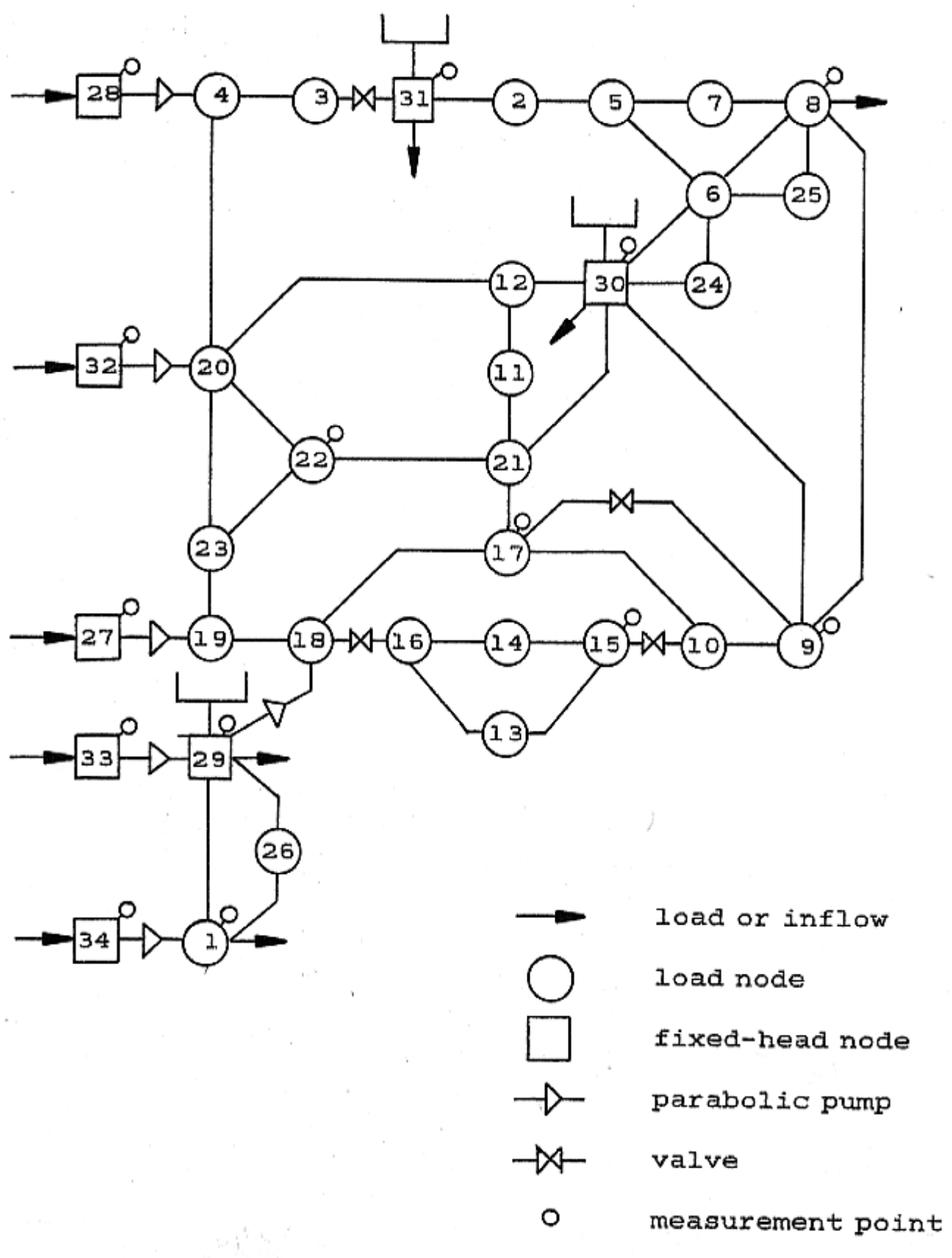

Figure 3. Water network specific to hilly areas.

There can be noticed 5 pumps which are located between nodes 34 and 1, 33 and 29, 27 and 19, 32 and 20, 28 and 4. Water is pumped in the water network by using these pumps. This justifies the negative heads at nodes 27, 28, 32, 33 and 34 in Table 1.

TABLE 1. TRUE AND OBSERVED FIXED-HEAD NODES

| **Fixed-head nodes [m]** | | |
|---|---|---|
| | ***True*** | ***Obs.*** |
| 27 | -15.1991 | -15.1991 |
| 28 | -33.4879 | -33.4978 |
| 29 | 31.7221 | 31.7221 |
| 30 | 43.5619 | 43.5819 |
| 31 | 44.1710 | 44.1703 |
| 32 | -46.3814 | -46.3814 |
| 33 | -36.5470 | -36.5478 |
| 34 | -12.1990 | -12.1963 |

The dissimilarities between the observed water distribution system state $\hat{x}$ and the true water distribution system state $x^t$ can be observed in Table 3. The reason is the insertion of the simulated measurement errors. It can be seen therefore how erroneous measurement data can influence deterministic water network state variables.

With the scope of surveying the quality of the ESM, the next study is implemented: first, the vector $\hat{x}$ is denoted the observed/optimal water distribution system state vector (i.e. columns 3 and 6 of Table 3). Afterwards, the observed measurement data $z^o$ is changed with $\Delta z$ in a random fashion and in agreement with the characteristics of the uncertainties of the water consumptions/pseudo-measurements and the fixed-head nodes.

For the randomly created measurement data $\Delta z$, the LS loop flows state estimator calculates a new water network state vector $\hat{x}_1$ and consequently a variation of the water network state vector $\Delta\hat{x}_1$ is determined as the distance between the observed water network state variable $\hat{x}$ and the new water network state vector $\hat{x}_1$. These numerical results are displayed also at Fig. 4.

TABLE 2. TRUE AND OBSERVED NODAL CONSUMPTIONS

| **Nodal consumptions [l/s]** | | | | | |
|---|---|---|---|---|---|
| ***Node*** | ***True*** | ***Obs.*** | ***Node*** | ***True*** | ***Obs.*** |
| 1 | 52.6 | 57.5 | 18 | 12.1 | 13.2 |
| 2 | 2.7 | 3.0 | 19 | 4.5 | 4.9 |
| 3 | 19.2 | 21 | 20 | 12.1 | 13.2 |
| 4 | 5.9 | 6.5 | 21 | 22.3 | 24.4 |
| 5 | 1.1 | 1.23 | 22 | 32.4 | 35.4 |
| 6 | 2.1 | 2.3 | 23 | 38.2 | 41.7 |
| 7 | 3.0 | 3.3 | 24 | 5.0 | 5.5 |
| 8 | 69.4 | 75.8 | 25 | 9.0 | 9.8 |
| 9 | 8.1 | 8.9 | 26 | 11.1 | 12.1 |
| 10 | 3.8 | 4.2 | 27 | 6.2 | 6.8 |
| 11 | 1.9 | 2.1 | 28 | 0 | 0 |
| 12 | 10.2 | 11.1 | 29 | 22.9 | 25 |
| 13 | 21.2 | 23.2 | 30 | 39.5 | 43.1 |
| 14 | 10.3 | 11.21 | 31 | 39.3 | 42.9 |
| 15 | 22.2 | 24.3 | 32 | 0 | 0 |
| 16 | 4.7 | 5.12 | 33 | 0 | 0 |
| 17 | 2.4 | 2.6 | 34 | 0 | 0 |

$$\Delta\hat{x}_1 = \hat{x}_1 - \hat{x} \qquad (20)$$

Secondly, the observed measurement vector $z^o$ is employed to calculate the ESM and represented as $S^e$ similar to (Eq. 17) as it is described next: for each $j$ greatest variation in the observed measurement vector $z_j$, is determined a changing in the observed water distribution system state vector for every element $x_i$ by using the LS loop flow state estimator; finally the component $s_{ij}$ of the matrix $S^e$ is determined as the operation $x_i / z_j$. Following this, the matrix $S^e$ is multiplied with the randomly variated measurement values $\Delta z$, and it results a variability of the water distribution system state vector $\Delta\hat{x}_2$ displayed in Fig. 4:

$$\Delta\hat{x}_2 = S^e \Delta z \qquad (21)$$

It is likely to happen that the greatest dissimilarity between the two curves $\Delta\hat{x}_1$ and $\Delta\hat{x}_2$ to be quite small. Actually the greatest dissimilarity is of only 0.05 m which is calculated between the nodal heads 7 and 8. Therefore it can be assumed for the time being that for a number of measurements and a range of errors connected with the respective measurements, the ESM could be a successful uncertainty method of calculating the water distribution system state variables for a variability in the input data measurements, which are comprising the nodal water consumptions and the nodal fixed-head pressures but without any real pipe flows or pressure measurements.

TABLE 3. TRUE AND OBSERVED WATER NETWORK STATE VARIABLES

| **True and observed water network state variables** | | | | | |
|---|---|---|---|---|---|
| ***Nodal pressures [m]*** | | | ***Nodal pressures [m]*** | | |
| ***Node*** | ***True*** | ***Obs.*** | ***Node*** | ***True*** | ***Obs.*** |
| 1 | 31.1852 | 31.0577 | 23 | 44.0663 | 43.9127 |
| 2 | 43.3886 | 43.2835 | 24 | 42.9028 | 42.7773 |
| 3 | 44.2289 | 44.1968 | 25 | 42.0751 | 41.7974 |
| 4 | 44.3191 | 44.2706 | 26 | 31.3306 | 31.2399 |
| 5 | 42.8133 | 42.6358 | 27 | -15.1991 | -15.1991 |
| 6 | 42.6765 | 42.5082 | 28 | -33.4879 | -33.4966 |
| 7 | 41.8478 | 41.5228 | 29 | 31.7221 | 31.7242 |
| 8 | 41.7190 | 41.3762 | 30 | 43.5619 | 43.5819 |
| 9 | 43.0165 | 42.8746 | 31 | 44.1710 | 44.1715 |
| 10 | 41.6933 | 41.1195 | 32 | -46.3814 | -46.3798 |
| 11 | 43.5925 | 43.5813 | 33 | -36.5470 | -36.5457 |
| 12 | 43.5845 | 43.5817 | 34 | -12.1990 | -12.1942 |
| 13 | 45.3550 | 45.2569 | ***Inflows [l/s]*** | | |
| 14 | 40.1661 | 39.2083 | ***Node*** | ***True*** | ***Obs.*** |
| 15 | 43.0940 | 39.1235 | 27 | 34.0 | 35.2 |
| 16 | 43.4858 | 43.0441 | 28 | 96.5 | 96.6 |
| 17 | 43.9047 | 43.7263 | 29 | 64.3 | 73.4 |
| 18 | 44.7605 | 44.5342 | 30 | 106.3 | 130.2 |
| 19 | 44.3638 | 44.1934 | 31 | 38.9 | 48.7 |
| 20 | 44.1362 | 44.0702 | 32 | 6.0 | 6.0 |
| 21 | 43.6560 | 43.6053 | 33 | 121.7 | 121.7 |
| 22 | 43.8080 | 43.7161 | 34 | 21.6 | 22.8 |

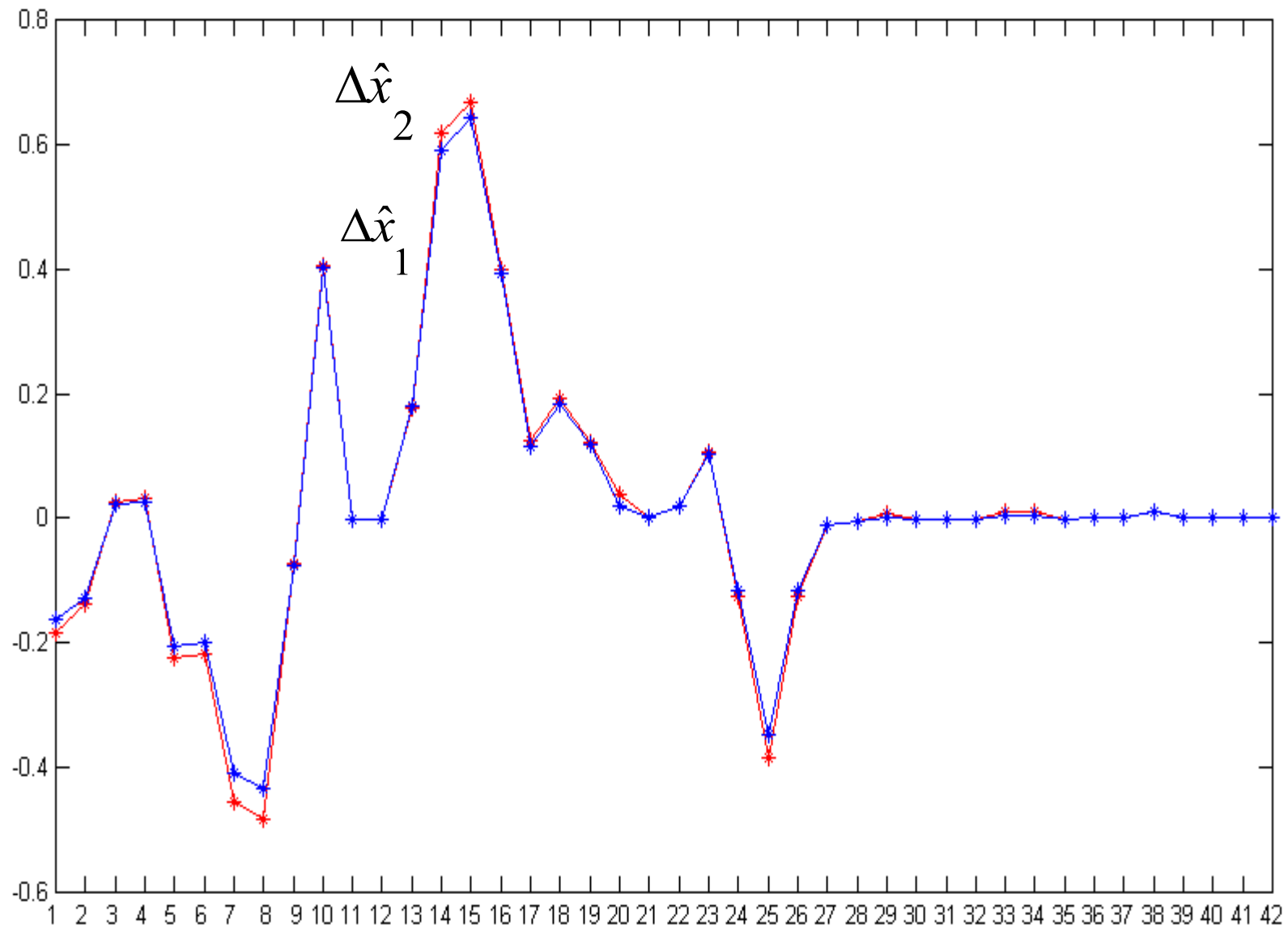

Figure 4. Variation of the water network state variables obtained with the ESM method $\Delta\hat{x}_2$ - Eq. 20 (above line) and the LS loop flows state estimator $\Delta\hat{x}_1$ - Eq. 21 (below line).

1-34: variation of nodal heads [m] at nodes 1-34.

35-42: variation of fixed-head nodes in/out flows [l/s] at nodes 27-34.

## V. CLA Based On ESM Method

Based on the newly derived matrix $S^e$, the augmentation method described in (Eq. 15) is implemented with the scope of determining the confidence limits for the water distribution system state variables. For the $i$-th water network state variable, determining the confidence limits is realized by maximizing the product between the $i$-th row of the ESM ($S^e$) and the vector $\Delta z$. The maximization process is realized for each row of $S^e$ calculated in the previous chapter. The confidence limits for the water network state variables (nodal heads and in/out flows) are displayed on the 4-th column of Table 4. The numerical results have been calculated for the variation of water network consumptions ±20% and the precision of fixed-head nodes ±0.01 [m].

TABLE 4. WATER NETWORK STATE VARIABLES AND CONFIDENCE LIMITS FOR THE 34-NODE WATER NETWORK

| Water Network State variable | Exact Water Network State Variable | LS loop flows state estimates | CLA obtained with ESM | LS nodal heads equations | CLA obtained with LS nodal heads equations |
|---|---|---|---|---|---|
| 1 | 31.1852 | 31.0577 | 0.3007 | 31.0566 | 0.3002 |
| 2 | 43.3886 | 43.2835 | 0.2557 | 43.2818 | 0.2604 |
| 3 | 44.2289 | 44.1968 | 0.0667 | 44.1960 | 0.0618 |
| 4 | 44.3191 | 44.2706 | 0.1013 | 44.2702 | 0.0965 |
| 5 | 42.8133 | 42.6358 | 0.4212 | 42.6336 | 0.4377 |
| 6 | 42.6765 | 42.5082 | 0.3971 | 42.5048 | 0.4166 |
| 7 | 41.8478 | 41.5228 | 0.7570 | 41.5214 | 0.8001 |
| 8 | 41.7190 | 41.3762 | 0.7925 | 41.3748 | 0.8441 |
| 9 | 43.0165 | 42.8746 | 0.3423 | 42.8282 | 0.3544 |
| 10 | 41.6933 | 41.0095 | 1.3561 | 40.9050 | 1.5754 |
| 11 | 43.5925 | 43.5813 | 0.0117 | 43.5815 | 0.0130 |
| 12 | 43.5845 | 43.5817 | 0.0033 | 43.5818 | 0.0035 |
| 13 | 45.3550 | 45.2569 | 1.3782 | 44.3555 | 1.9953 |
| 14 | 40.1661 | 39.3083 | 2.2626 | 39.5726 | 2.8137 |
| 15 | 43.0940 | 39.2235 | 2.2657 | 39.4799 | 2.8731 |
| 16 | 43.4858 | 43.0441 | 1.0413 | 43.1425 | 0.9928 |
| 17 | 43.9047 | 43.7263 | 0.4081 | 43.6956 | 0.3504 |
| 18 | 44.7605 | 44.5342 | 0.5268 | 44.5576 | 0.4893 |
| 19 | 44.3638 | 44.1934 | 0.3900 | 44.2076 | 0.3748 |
| 20 | 44.1362 | 44.0702 | 0.1410 | 44.0704 | 0.1360 |
| 21 | 43.6560 | 43.6053 | 0.1006 | 43.6056 | 0.0938 |
| 22 | 43.8080 | 43.7161 | 0.1993 | 43.7192 | 0.1899 |
| 23 | 44.0663 | 43.9127 | 0.3484 | 43.9223 | 0.3344 |
| 24 | 42.9028 | 42.7773 | 0.3027 | 42.7718 | 0.3119 |
| 25 | 42.0751 | 41.7974 | 0.6471 | 41.7951 | 0.6847 |
| 26 | 31.3306 | 31.2399 | 0.2259 | 31.2384 | 0.2096 |
| 27 | -15.1991 | -15.1991 | 0.0000 | -15.1991 | 0.0185 |
| 28 | -33.4879 | -33.4966 | 0.0151 | -33.4978 | 0.0191 |
| 29 | 31.7221 | 31.7242 | 0.0196 | 31.7221 | 0.0116 |
| 30 | 43.5619 | 43.5819 | 0.0004 | 43.5819 | 0.0102 |
| 31 | 44.1710 | 44.1715 | 0.0151 | 44.1703 | 0.0183 |
| 32 | -46.3814 | -46.3798 | 0.0151 | -46.3810 | 0.0197 |
| 33 | -36.5470 | -36.5457 | 0.0161 | -36.5478 | 0.0142 |
| 34 | -12.1990 | -12.1942 | 0.0135 | -12.1963 | 0.0159 |
| 35 | 34.0 | 35.2 | 3.1 | 35.6 | 3.7 |
| 36 | 96.5 | 96.6 | 0.1 | 97.0 | 0.1 |
| 37 | 64.3 | 73.4 | 21.4 | 73.4 | 21.5 |
| 38 | 106.3 | 130.2 | 56.9 | 130.1 | 56.2 |
| 39 | 38.9 | 48.7 | 22.8 | 48.3 | 22.0 |
| 40 | 6.0 | 6.0 | 0 | 6.0 | 0 |
| 41 | 121.7 | 121.7 | 0 | 121.7 | 0 |
| 42 | 21.6 | 22.8 | 2.6 | 22.8 | 2.6 |

1-34: nodal heads [m] at nodes 1-34;

35-42: fixed-head nodes in/out flows [l/s] at nodes 27-34.

On the 5-th and 6-th column of Table 4 are displayed the water distribution system state variables and the confidence limits (CLA) determined by using the Jacobian matrix from the LS nodal heads state estimator (i.e. sensitivity matrix method from the LS nodal heads state estimator). As already explained in [9][25] the sensitivity matrix method from the LS nodal heads state estimator determines more precise confidence limits (i.e. good confidence limits) for the water distribution system state variables as opposed to other algorithms such as an ellipsoid procedure [25][112]. It also determines similar confidence limits with a linear programming method and the Monte Carlo algorithm [25]. Therefore, the sensitivity matrix method realized by the LS nodal heads state estimator is employed for the assessment of the novel algorithms presented herein and which are built on top of the LS loop flows state estimator.

Without using real measurements, the confidence limits determined from the ESM method on one hand and the pseudo-inverse of the Jacobian matrix from the LS nodal heads state estimator on the second hand, demonstrated some similarity: the average difference between the 2 techniques (i.e.

difference between column 4 and 6 of Table 4) over all the nodal heads is 0.0725 [m] and 0.2715 [l/s] for the in/out flows.

Therefore, the matrix $S^e$ shows the characteristics of the inverse of the Jacobian matrix from the LS nodal heads simulator algorithm and it can be employed with the scope of calculating the water distribution system state variables and the confidence limits.

However, this works in the absence of pressure and flow measurements in the numerical simulations or the state estimations as the ones displayed at Table 4. However, the use of additional pressure and pipe flow measurements will determine the appearance of non-linearities in the determination of the $s_{ij}$ component of the matrix $S^e$ which will affect the respective operations of type $x_i / z_j$ which will not be able to express anymore the relationship between the calculated water distribution system state variable $x_i$ (i.e. the observed water network state vector $\hat{x}$) because of the maximum $j$ variability in the observed measurement $z_j$ (i.e. the observed measurement vector $z^o$).

Moreover, although the ESM method and $S^e$ matrix are effective in providing realistic water network state vectors and confidence limits when there are no pipe flows and nodal pressures measurements for any operational point, they require as many simulations as the number of real measurements and pseudo-measurements. Therefore the computational complexity tends to be a major drawback because even for a small-sized water system, the number of feasible measurements (i.e. nodal consumptions and fixed-head nodes) is great, rendering this approach difficult to use in on-line decision support system.

In view of these limitations, an alternative method has been developed which solves the linearized model of the water network for the maximum of errors in the estimated measurement vector $\hat{z}$ calculated from the observed water network state vector $\hat{x}$, which is obtained after running the LS loop flows state estimator for the observed measurement vector $z^o$.

## VI. Error Maximization Method

The novel EM algorithm corresponds that to a Maximum amount (M) of the uncertainties/Errors (E) in the input measurement data of the water network, it can be obtained appropriate confidence limits for the water distribution system state variables and therefore the EM term was coined for this method.

The experimental matrix $S^e$ obtained for the water distribution system state vector $\hat{x}$ and the upper $z^u$ and the lower confidence limits $z^l$ of the observed measurement vector $z^o$ were employed in the ESM technique as displayed in Fig. 5 and in the same time with the LS loop flows state estimator and the observed measurement vector $z^o$ and it finally outputted the 'observed' water distribution system state vector $\hat{x}$. Therefore, an experimental matrix was determined by employing repetitively the LS loop flows state estimator for a number of times equal to the number of water nodal consumptions and the fixed-head nodes but without involving any nodal heads or pipe flows measurements.

Fig. 5 compares the ESM technique with the novel EM technique which takes into consideration the maximum variations in the water nodal consumptions and the inexactness of real-life meters (i.e. nodal heads and pipe flows measurements) for the estimated measurement vector $\hat{z}$ (i.e. determined based on the 'observed' water distribution system state vector $\hat{x}$ ) and without employing the observed measurement vector $z^o$.

Then, the upper or the lower measurement bounds [ $\hat{z}^l$ $\hat{z}^u$ ] of the estimated measurement vector $\hat{z}$, which is changed with its measurement preciseness, are employed in return by the LS loop flows state estimator. It derives a new water distribution system state vector $x^1$ which is employed for calculating the confidence limits on the water distribution system state variables (nodal heads, in/out flows) with the equation (22). Furthermore, in Figure 6 the block diagram can deliver two quantities, which are $x^{1u}$ and $x^{11}$, one corresponding to $\hat{z}^l$ and second to $\hat{z}^u$, by using two times the LS loop flows state estimator. Based on the two results of the state estimation procedure, $x^{1u}$ and $x^{11}$, one can go further and determined the confidence limits which would be $xcl_i{}^u$ and also $xcl_i{}^l$ with equation (22). However for all the numerical state estimations/simulations implemented for two water networks there were not noticed any statistical dissimilarities between $xcl_i{}^u$ and $xcl_i{}^l$ by using a mathematical t-test. So it was inferred that any of the two $\hat{z}^l$ or $\hat{z}^u$ vectors can be employed and one group of confidence limits $xcl_i$ can be determined, or for investigatory aims, both groups of confidence limits could be determined and shown to the human operator $xcl_i{}^u$ and $xcl_i{}^l$, and this would necessitate using for the second time the LS loop flows state estimator.

Henceforth, it will be accepted that since there were not noticed any statistical dissimilarities between ${xcl_i}^u$ and ${xcl_i}^l$ in the numerical simulations/state estimations implemented, then only one group of confidence limits will be shown and named commonly with $xcl_i$:

$$xcl_i = abs\,(\,x^1 - \hat{x}\,) \qquad (22)$$

where $xcl_i$ is the confidence limit on the $i$-th water distribution system state variable, $\hat{x}$ is the water distribution system state vector calculated for the observed measurement vector $z^o$ with the LS loop flows state estimator, $x^1$ is the water distribution system state vector calculated with the LS loop flows state estimator for the highest amount of errors (i.e. $\hat{z}^l$ or $\hat{z}^u$) in the estimated measurement vector $\hat{z}$, *abs* is the absolute quantity.

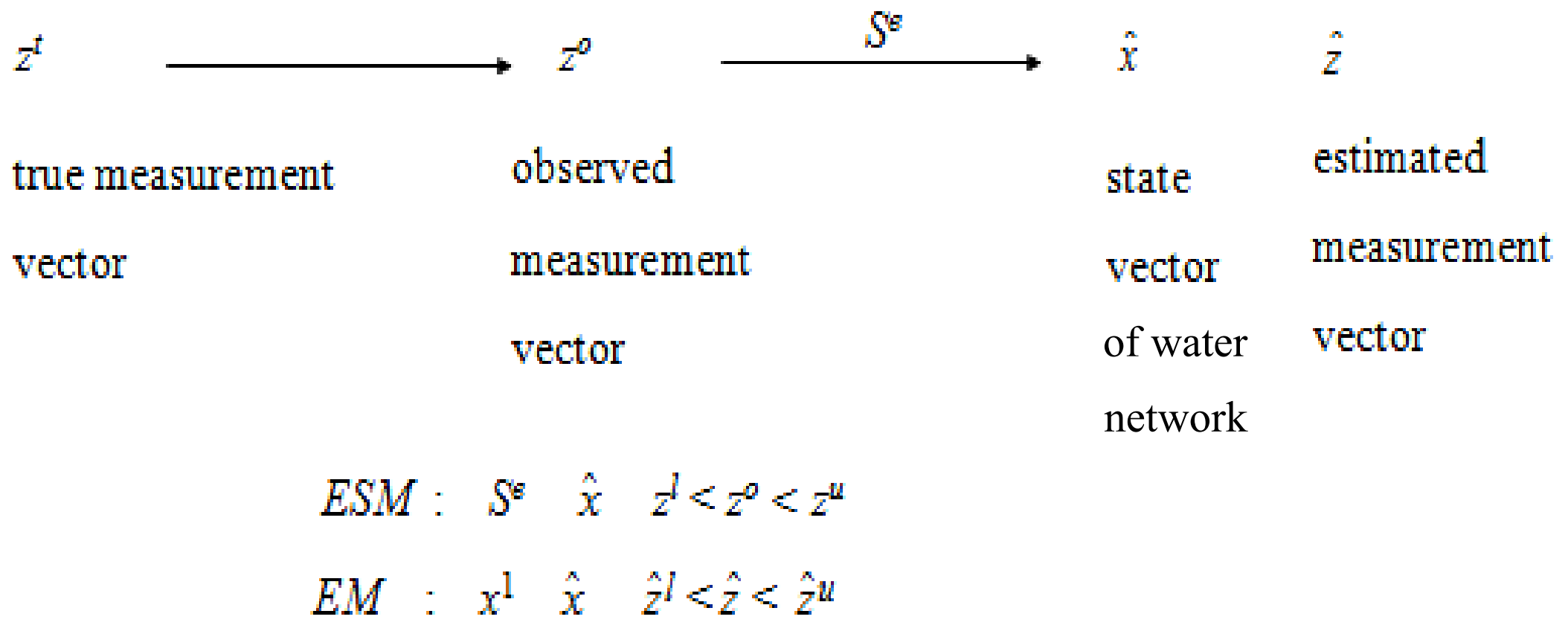

Figure 5. Brief comparison between ESM and EM methods: ESM relies on $S^e$ (computational burden to obtain) while EM on relies on vector $x^1$ (not computational burden to obtain).

Fig. 6 shows schematizes the EM method.

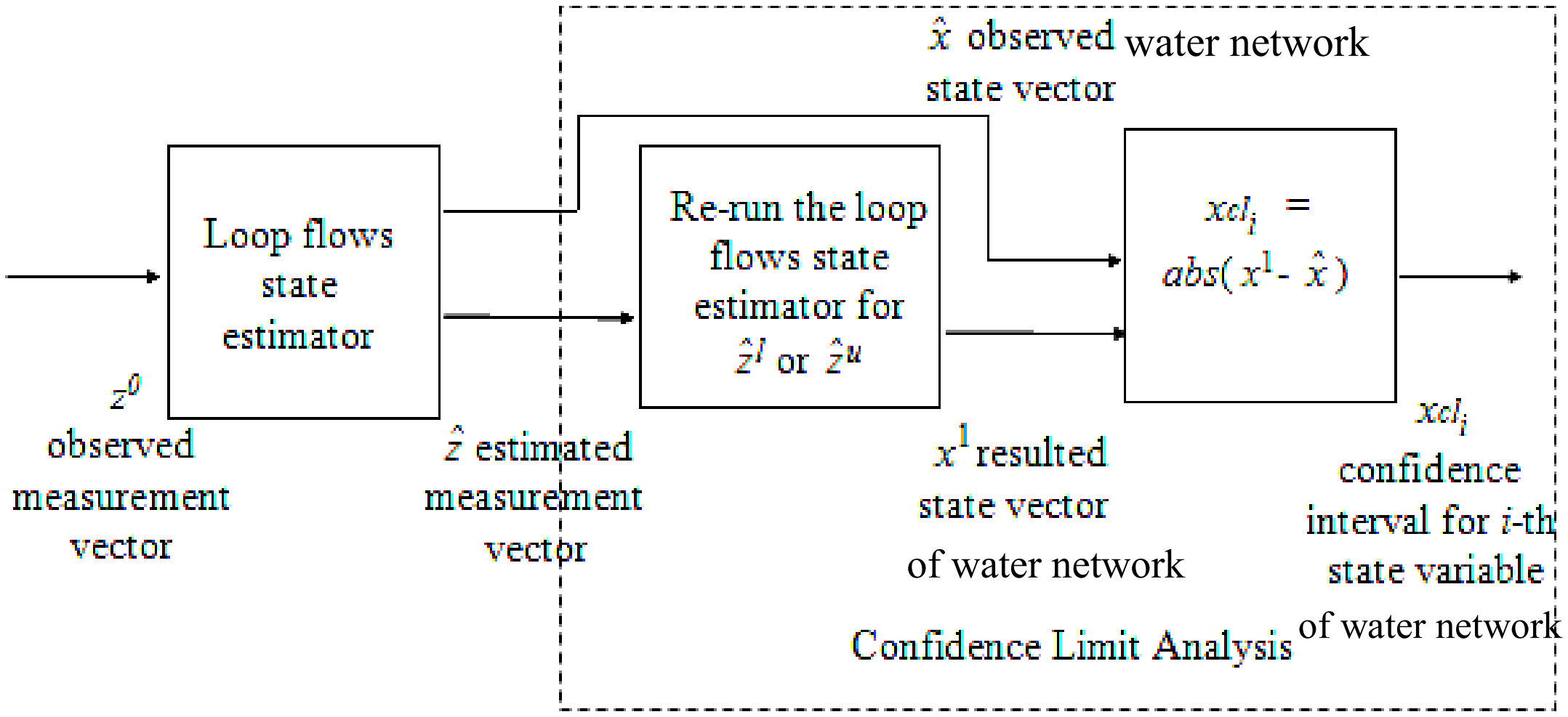

Figure 6. CLA based on EM method.

The EM technique can be easily described as follows:

**EM Algorithm.** The Error Maximization technique for uncertainty study of water network measurement data:

1. By using the observed measurement vector $z^o$, the LS loop flows state estimator is employed with any mixture of real life pipe flows or nodal pressure measurements to calculate the observed water distribution system state vector $\hat{x}$ which would include all the nodal pressures, pipe flows and in/out flows.
2. Based on the observed water distribution system state vector $\hat{x}$, it is determined the estimated measurement vector $\hat{z}$ which includes the water nodal demands, fixed nodal heads and any mixture of real life nodal heads and pipe flows.
3. The upper or the lower confidence limits [ $\hat{z}^l$ $\hat{z}^u$ ] of the estimated measurement vector $\hat{z}$ changed with the maximum measurement preciseness, are employed again by the LS loop flows state estimator and it is calculated a new water distribution system state vector $x^1$ (i.e. the 'estimated' water distribution system state vector).
4. The calculated water distribution system state vector $x^1$ is employed for obtaining the confidence limits $xcl$ on the water distribution system state variables (nodal pressures, in/out flows) with the Eq. (22) written again below so that to be clear:

$$xcl_i = abs\,(\,x^1 - \hat{x}\,)$$

Determining the confidence limits with the EM technique is characteristic to the LS loop flows state estimator and it is not possible to be implemented with the LS nodal heads state estimator which is not using the loop corrective flows and the variation of nodal demands as it does the LS nodal heads state estimator.

It is paramount to investigate more and to understand the hydraulic background of the EM technique which sustains also why the EM technique is so successful with the LS loop flows state estimator. As an illustration there are presented two completely different cases in which the determination of the confidence limits with the EM technique can be realized.

In the first illustration (I) the LS loop flows state estimator changes the in/out flows at the fixed-nodal heads so that to cope with the continuity equations and the total of the estimated nodal consumptions which are part of the estimated measurement vector $\hat{z}$ determined with the LS loop flows state estimator. This has the particular characteristic that in the situation that the water nodal demands from vector $\hat{z}$ are shifted to their lower $\hat{z}^l$ or upper bounds $\hat{z}^u$ then the continuity equations in the water distribution system will still be coped with by the in/out flows at the fixed-nodal heads or the main root node and determined by the Newton-Raphson numerical technique. In this situation, the fixed-nodal heads are the measurement information and are employed to build pseudo-loops connected to the main source node. The main source node is the root node which is used to construct the spanning tree, with the scope of producing the topological incidence matrixes required by the loop simulator algorithms and the LS loop flows state estimator [1][8][15][16].

In the second illustration (II), the in/out flows at the fixed-nodal heads correspond to the measurement information. The in/out flows at the fixed-nodal heads are given and maintained the same during the Newton-Raphson numerical technique. The continuity equations in the water distribution system are still complied with in the LS loop flows state estimator as there exists the main source node where the in/out flow can modify accordingly. The two different situations from above are exhibited in the Fig. 7. It exhibits the spanning tree of a small water distribution system. The black square corresponds to the main source node where the in/out flow is not kept the same during the Newton-Raphson numerical technique. The circles are the water distribution system nodes while the empty squares are the fixed-nodal heads. A nodal head measurement is shown with the identification *P*. The measurement preciseness is shown in the figure by an arrow with two wedge-shape ends. The white arrows identify the preciseness of the water nodal demands or the inflows/outflows at the fixed-nodal heads. The black arrows identify the greatest possible variation of the fixed-nodal heads and the nodal heads measurements.

Fig. 7a exhibits the first illustration where the fixed-head nodes belong to the measurement information and their nodal heads values are taken as being known and employed to build the pseudo-loops (i.e. the dashed lines in Fig.7). The lower $\hat{z}^l$ or the upper limits $\hat{z}^u$ of the nodal demands are used and the LS loop flows state estimator is employed to calculate the confidence limits. The fixed-nodal heads and the real metering information can be also changed in agreement to their highest values of the measurement preciseness. In Fig. 7b is exhibited the second illustration where the estimates of the fixed-head nodes are not determined yet and instead the inflows/outflows at fixed-head nodes are used as measurements and maintained constant during the Newton-Raphson technique. If all the measurement information is taken as the lower or upper bound then the mass balance equation of the water distribution system is coped with by only the main source node in/out flow.

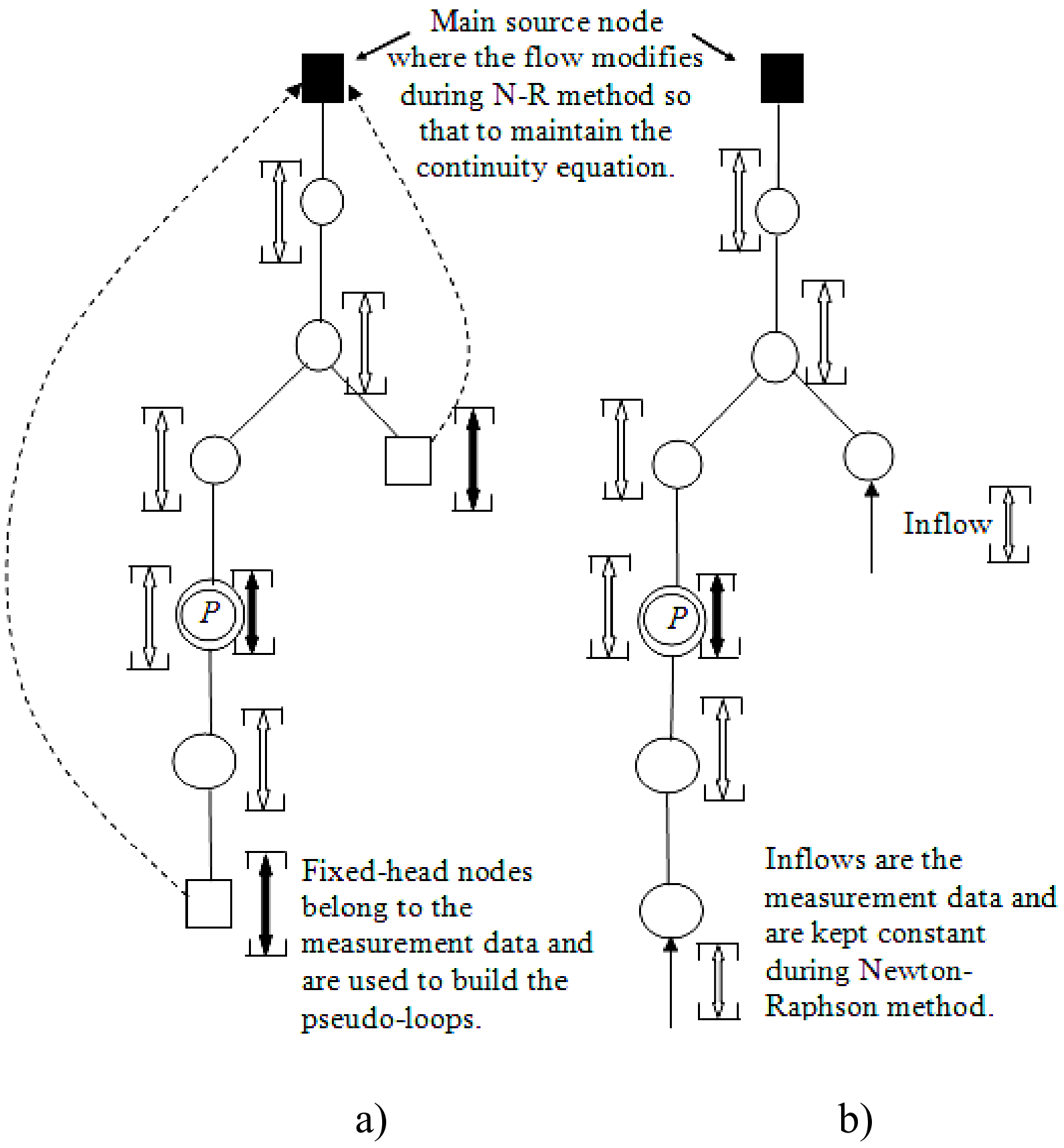

Figure 7. Two examples which can be met in the LS loop flows state estimator and the EM method: a) Fixed-head nodes used to form the pseudo-loops and form the measurement data (I), b) Inflows are measurement data (II).

It is obvious that the EM technique is suitable to the LS loop flows state estimator. This is because of the existence of the main source node where the in/out flow can not be kept at a fixed value and it modifies in agreement to the water nodal consumptions and the other pipe flows and nodal heads measurements from the water distribution system during the Newton-Raphson numerical technique.

This is different from the LS nodal heads state estimator in which if the water distribution system is state estimated while all the measurement information is taken equal to the lower or the upper error limit and in agreement to the measurements accuracy, then, the newly calculated LS nodal heads state estimated vector would not be possible to be used for determining the confidence limits because of the mathematical definition of the LS nodal heads state estimator [6]. Instead, the Jacobian matrix determined with the LS nodal heads state estimator can be employed for the determination of the confidence limits and this forms the sensitivity matrix method in the context of the LS nodal heads state estimator [9][10][21].

Therefore, the confidence limits of the nodal heads and pipe flows because of the inexactness in the nodal consumptions and other measurement data are determined with the EM technique by running one more time the LS loop flows state estimator.

This situation exists as the inflow/outflow at the main source node (i.e. from where the spanning tree is constructed) as well as the other fixed-head nodes (Figure 7a) can not be kept fixed and they modify by taking into account the estimated nodal demands and the other (few) measurements from the water distribution system. Therefore, the confidence limits of the nodal heads and pipe flows can be determined with equation (22).

With the aim of investigating the efficacy of the novel EM technique, a comparison of the confidence limits outputted by the ESM technique and the EM technique is exhibited at Table 5 when there are not employed any nodal heads or pipe flows measurements. The confidence limits are determined for the observed measurement vector exhibited on the columns 3 and 6 of Table 2 and column 3 of Table 1 for the water distribution system exhibited in Fig 3.

TABLE 5. CONFIDENCE LIMITS OBTAINED WITH THE ESM AND EM METHODS

| **Water Network State variables** | **Exact Water Network state** | **Observed Water Network State variables** | **C.L. with ESM method** | **C.L. with EM method** |
|---|---|---|---|---|
| 1 | 31.1852 | 31.0577 | 0.3007 | 0.2893 |
| 2 | 43.3886 | 43.2835 | 0.2557 | 0.2381 |
| 3 | 44.2289 | 44.1968 | 0.0867 | 0.1056 |
| 4 | 44.3191 | 44.2706 | 0.1213 | 0.1505 |
| 5 | 42.8133 | 42.6358 | 0.4212 | 0.4010 |
| 6 | 42.6765 | 42.5082 | 0.3971 | 0.3785 |
| 7 | 41.8478 | 41.5228 | 0.7570 | 0.7302 |
| 8 | 41.7190 | 41.3762 | 0.7925 | 0.7700 |
| 9 | 43.0165 | 42.8746 | 0.3423 | 0.3233 |
| 10 | 41.6933 | 41.1195 | 1.3561 | 1.3034 |
| 11 | 43.5925 | 43.5813 | 0.0117 | 0.0511 |
| 12 | 43.5845 | 43.5817 | 0.0033 | 0.0152 |
| 13 | 45.3550 | 45.2569 | 1.0782 | 0.9456 |
| 14 | 40.1661 | 39.2083 | 2.2626 | 2.1782 |
| 15 | 43.0940 | 39.1235 | 2.2657 | 2.1947 |
| 16 | 43.4858 | 43.0441 | 1.0413 | 1.0164 |
| 17 | 43.9047 | 43.7263 | 0.4081 | 0.4438 |
| 18 | 44.7605 | 44.5342 | 0.5268 | 0.5347 |
| 19 | 44.3638 | 44.1934 | 0.3900 | 0.4137 |
| 20 | 44.1362 | 44.0702 | 0.1410 | 0.1951 |
| 21 | 43.6560 | 43.6053 | 0.1006 | 0.1476 |
| 22 | 43.8080 | 43.7161 | 0.1993 | 0.2445 |
| 23 | 44.0663 | 43.9127 | 0.3484 | 0.3746 |
| 24 | 42.9028 | 42.7773 | 0.3027 | 0.2826 |
| 25 | 42.0751 | 41.7974 | 0.6471 | 0.6240 |
| 26 | 31.3306 | 31.2399 | 0.2259 | 0.1880 |
| 27 | -15.1991 | -15.1991 | 0.0000 | 0.0000 |
| 28 | -33.4879 | -33.4966 | 0.0151 | 0.0112 |
| 29 | 31.7221 | 31.7242 | 0.0196 | 0.0112 |
| 30 | 43.5619 | 43.5819 | 0.0004 | 0.0000 |
| 31 | 44.1710 | 44.1715 | 0.0151 | 0.0141 |
| 32 | -46.3814 | -46.3798 | 0.0151 | 0.0139 |
| 33 | -36.5470 | -36.5457 | 0.0201 | 0.0121 |
| 34 | -12.1990 | -12.1942 | 0.0199 | 0.0141 |
| 35 | 34.0 | 35.2 | 3.1 | 2.9 |
| 36 | 96.5 | 96.6 | 0.1 | 0.2 |
| 37 | 64.3 | 73.4 | 21.4 | 21.9 |
| 38 | 106.3 | 130.2 | 56.9 | 55.5 |
| 39 | 38.9 | 48.7 | 22.8 | 23.9 |
| 40 | 6 | 6 | 0 | 0 |
| 41 | 121.7 | 121.7 | 0 | 0 |
| 42 | 21.6 | 22.8 | 2.6 | 2.6 |

1-34: nodal heads [m] at nodes 1-34;

35-42: fixed-head nodes in/out flows [l/s] at nodes 27-34.

The confidence limits determined with the ESM and EM techniques are very similar because the average difference between the 2 techniques over all the nodal heads is 0.0280 [m] and 0.4125 [l/s] for the in/out flows. For example, an inflow at node 30 is of 106.3 [l/s] with a confidence limit of 56.9 [l/s].

The average difference for the confidence limits between the EM technique and the sensitivity matrix technique based on the LS nodal heads equations exhibited on the 6-th column of Table 4 is 0.1 [m] for the nodal heads and 0.4875 [l/s] for the in/out flows. It can be inferred that the EM technique is qualitatively the same as both the ESM technique and the sensitivity matrix method based on the LS nodal heads state estimator, when no real pipe flows and nodal heads measurements are taken into consideration.

The computational time required by the ESM technique, which was over 15 seconds, it was also by far higher than the less of 0.5 second resulted from using the EM technique, which needed also only to run once the LS loop flows state estimator.

This is because of the computational overload required to obtain the experimental sensitivity matrix $S^e$ (ESM). In order to calculate the ESM, the LS loop flows state estimator was used successively for each variability of the pseudo-measurements and the fixed-head nodal pressures.

An additional number of measurements will necessitate an equal number of extra state estimations of the water distribution system which will add to the computational time required to determine the ESM especially for large water distribution systems.

The EM technique when using real pipe flows and nodal pressures measurements is described below.

TABLE 6. CONFIDENCE LIMITS OBTAINED WITH THE EM METHOD WHEN REAL METERS ARE PRESENT

| **Water Network State variables** | **Values of Water Network State variables** | **C.L. (case 1)** | **C.L. (case 2)** | **C.L. (case 3)** |
|---|---|---|---|---|
| 1 | 31.0577 | 0.2893 | 0.2801 | 0.2593 |
| 2 | 43.2835 | 0.2381 | 0.2448 | 0.2234 |
| 3 | 44.1968 | 0.1056 | 0.0842 | 0.0455 |
| 4 | 44.2706 | 0.1505 | 0.1123 | 0.0636 |
| 5 | 42.6358 | 0.4010 | 0.4042 | 0.3824 |
| 6 | 42.5082 | 0.3785 | 0.3832 | 0.3616 |
| 7 | 41.5228 | 0.7302 | 0.7254 | 0.7030 |
| 8 | 41.3762 | 0.7700 | 0.7641 | 0.7416 |
| 9 | 42.8746 | 0.3233 | 0.1553 | 0.1201 |
| 10 | 41.1195 | 1.3034 | 0.7980 | 0.8301 |
| 11 | 43.5813 | 0.0511 | 0.0401 | 0.0098 |
| 12 | 43.5817 | 0.0152 | 0.0199 | 0.0028 |
| 13 | 45.2569 | 0.9456 | 1.0041 | 1.4464 |
| **14** | **39.2083** | **2.1782** | **1.1753** | **1.1761** |
| **15** | **39.1235** | **2.1947** | **1.1631** | **1.1638** |
| **16** | **43.0441** | **1.0164** | **0.5342** | **0.4530** |
| 17 | 43.7263 | 0.4438 | 0.1043 | 0.0991 |
| 18 | 44.5342 | 0.5347 | 0.2145 | 0.0141 |
| 19 | 44.1934 | 0.4137 | 0.2307 | 0.0485 |
| 20 | 44.0702 | 0.1951 | 0.1397 | 0.0789 |
| 21 | 43.6053 | 0.1476 | 0.1026 | 0.0484 |
| **22** | **43.7161** | **0.2445** | **0.1737** | **0.0931** |
| **23** | **43.9127** | **0.3746** | **0.2432** | **0.0869** |
| 24 | 42.7773 | 0.2826 | 0.2916 | 0.2703 |
| 25 | 41.7974 | 0.6240 | 0.6224 | 0.6002 |
| 26 | 31.2399 | 0.1880 | 0.1988 | 0.1780 |
| 27 | -15.1991 | 0.0000 | 0.0100 | 0.0100 |
| 28 | -33.4966 | 0.0112 | 0.0095 | 0.0112 |
| 29 | 31.7242 | 0.0112 | 0.0087 | 0.0122 |
| 30 | 43.5819 | 0.0000 | 0.0100 | 0.0100 |
| 31 | 44.1715 | 0.0141 | 0.0095 | 0.0113 |
| 32 | -46.3798 | 0.0139 | 0.0095 | 0.0112 |
| 33 | -36.5457 | 0.0121 | 0.0087 | 0.0122 |
| 34 | -12.1942 | 0.0141 | 0.0087 | 0.0122 |
| 35 | 35.2 | 2.9 | 1.5 | 0.4 |
| 36 | 96.6 | 0.2 | 0.1 | 0.1 |
| 37 | 73.4 | 21.9 | 18.4 | 16.5 |
| 38 | 130.2 | 55.5 | 46.1 | 41.1 |
| 39 | 48.7 | 23.9 | 20.8 | 18.8 |
| 40 | 6 | 0 | 0 | 0 |
| 41 | 121.7 | 0 | 0 | 0 |
| 42 | 22.8 | 2.6 | 2.6 | 2.6 |

1-34: nodal heads [m] at nodes 1-34;

35-42: fixed-head nodes in/out flows [l/s] at nodes 27-34.

On the 3-rd column of Table 6 there are exhibited the confidence limits determined with the EM technique for the variability of pseudo-measurements and preciseness of the fixed-head nodes but no other real metering devices are used. In order to avoid complicating the study, the estimated measurement vector is taken similar to the observed measurement data from Tables 1 and Table 2.

By determining the confidence limits, it is attained a clear insight about how far from the real functioning point of the water distribution system, the estimated water distribution system values could be in the worst situation. Having determined the water distribution system state variables as close as possible to the real state, this is the same as determining as much as possible very tight confidence limits. This is obtained by placing more exact measuring devices (pipe flow meters or nodal heads measurements) into the water network.

The EM technique can use nodal heads and pipe flow measurements as well. A nodal head measurement is located at node 14 of the water distribution system from Fig. 3 that translates into the row 4 of Table 6 (C.L case 2) . The preciseness of the nodal head measuring device is $\pm0.03\%$.

Therefore, on the 4-th column of Table 6 (CL case 2) are exhibited the confidence limits when a nodal head measurement is located at node 14.

As it is expected, tighter confidence limits of -1 (m) are achieved, not only in the node where the nodal pressure measurement was located that is node 14, displayed also with bold line, but also in the adjacent nodes to node 14 that are the nodes 15 and 16, displayed also with bold line in Table 6.

The same theory is considered for the pipe flow measurements (C.L. case 3): a pipe flow measurement with the preciseness of $\pm0.02\%$ is placed between nodes 22 and 23. On the 5-th column of Table 6 an enhancement, that is smaller confidence limits, can be seen in the area where the pipe flow measurement was located, that is nodes 22 and 23.

In conclusion, with a higher amount of exact measurements, the confidence in the state estimation becomes grater.

However, by placing a new measurement it can happen to bring in a new source of inexactness which is defined by the tolerance of the measuring device.

Therefore, when considering a new measurement for the *i*-th water distribution system state variable, this can produce the tightening effect on the confidence limits of the respective water distribution system state variable only if the error produced by the inexactness of the measuring device is smaller than the confidence limit determined based on the existing group of meters. If the previous condition is met, then the confidence limits for the nodal pressures and inflows/outflows become tighter as well.

With the scope of underlining the EM technique, there are exhibited more simulation examples with the EM technique for determining the confidence limits for the realistic size water distribution system from Fig. 8. The data characterizing this water distribution system is exhibited in the Appendix 2.

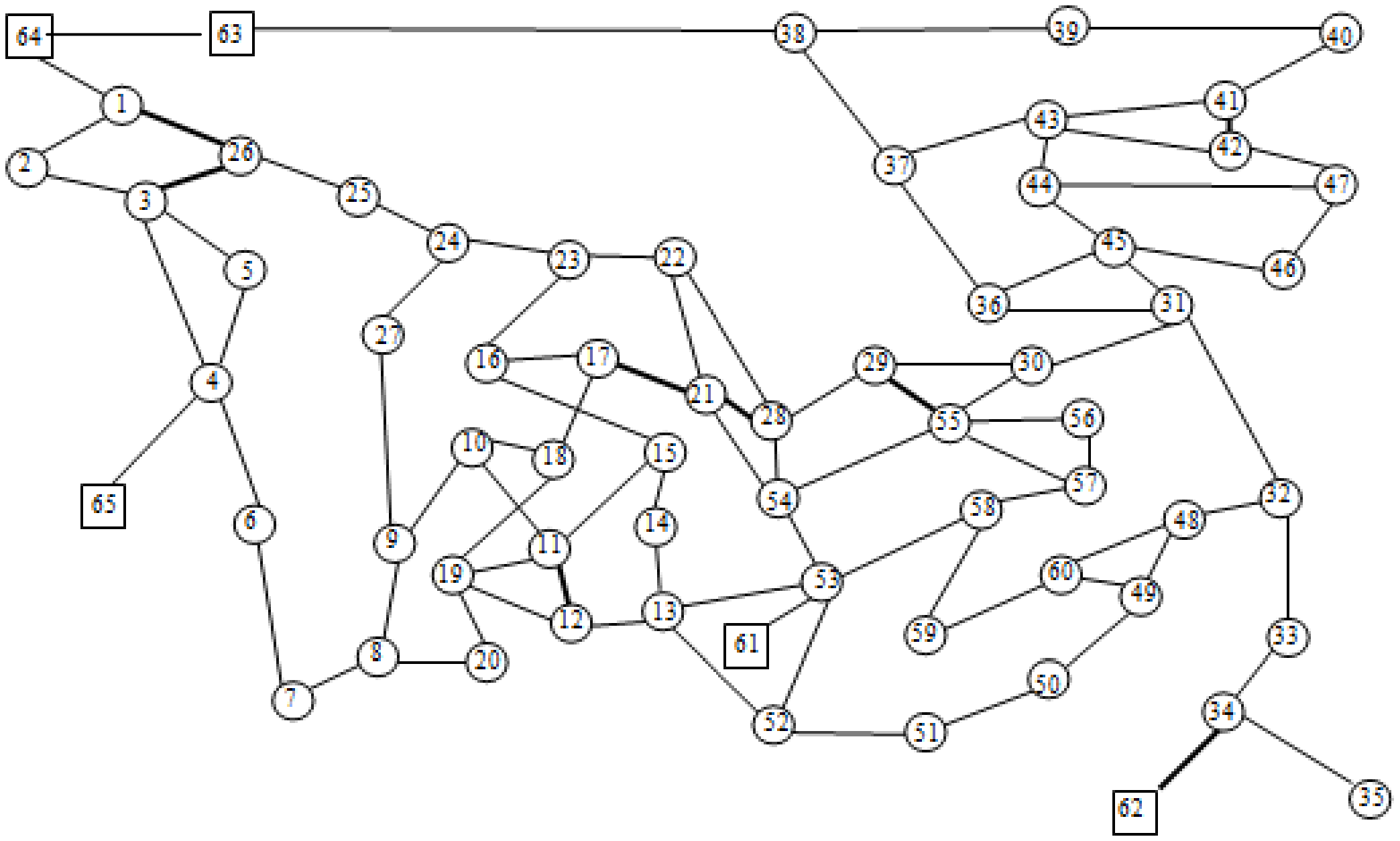

Figure 8. Water network

On the 2nd column of Table 7 are exhibited the water distribution system state variables for the water distribution system parameters and measurement data exhibited in Table 10 and Table 11 from Appendix 2.

Following this, for the variation of nodal pseudo-measurements of +20%, the confidence limits are determined for the measurement data exhibited in Table 11 of Appendix 2.

The determined confidence limits on the nodal pressures and the water distribution system inflows are displayed on the 3rd column of Table 7 (CL case 1) with the EM technique for the maximum variation in the measurement data $\hat{z}^{u}$ (i.e. +20%) consisting of the nodal pseudo-measurements.

The variation of the fixed-head nodes (i.e. nodes 61, 62, 63, 64 and 65) is taken as zero, which corresponds to the fact that the fixed-head nodes are considered to be exact and consequently their confidence limits exhibited on the same column 3 of Table 7 are also null.

The confidence limits exhibited on the 4th column of Table 7 (CL case 2) are determined for the maximum variation in the measurement data $\hat{z}^{u}$ including the nodal pseudo-measurements modified with the maximum error of +20%. Also, an exact nodal head measurement is placed at node 5. It can be observed that the confidence limits exhibited on the 4th column of Table 7 are smaller in size, as opposed to the confidence limits exhibited on the 3rd column of Table 7 and the reasons are the placement of the exact nodal heads measurement at node 5. Also the confidence limits for the nodal head measurement, that is the node 5, are also zero because the respective nodal head measurement is taken as being exact and not affected by inaccuracies.

The confidence limits exhibited on the 5th column of Table 7 (CL case 3) are determined for a similar maximum variation in the measurement data $\hat{z}^u$ which includes the nodal pseudo-measurements modified with the maximum amount of +20%.

Now, an exact nodal head measurement is placed at node 5 and also three exact pipe flow measurements in pipes 9-10, 59-60 and 39-40. The confidence limits exhibited on the 5th column of Table 7 point out to smaller confidence limits as opposed to the confidence limits displayed on the 4th column of Table 7 and the reason is the use of the 3 additional exact pipe flows measurements.

For the pair of nodes 9-10, 59-60 and 39-40 it can be seen tighter confidence limits for the respective nodes on the 5th column in comparison to the 4th column.

TABLE 7. CONFIDENCE LIMITS OBTAINED WITH THE EM METHOD WHEN REAL METERS ARE PRESENT ON REALISTIC WATER NETWORK

| **Water Network State variables** | **Values of Water Network State variables** | **C.L. (case 1)** | **C.L. (case 2)** | **C.L. (case 3)** |
|---|---|---|---|---|
| 1 | 143.49 | 0.2363 | 0.2133 | 0.1698 |
| 2 | 142.83 | 0.2675 | 0.2321 | 0.1842 |
| 3 | 141.73 | 0.2580 | 0.2092 | 0.1818 |
| 4 | 141.63 | 0.1699 | 0.1358 | 0.1114 |
| **5** | **141.25** | **0.3693** | **0** | **0** |
| 6 | 141.15 | 0.4921 | 0.4399 | 0.3489 |
| 7 | 140.81 | 0.7308 | 0.6672 | 0.5218 |
| 8 | 140.63 | 0.8575 | 0.7891 | 0.6119 |
| 9 | 140.39 | 0.9908 | **0.9184** | **0.7039** |
| 10 | 140.35 | 1.0112 | **0.9381** | **0.7114** |
| 11 | 140.28 | 1.0555 | 0.9813 | 0.7372 |
| 12 | 140.27 | 1.0580 | 0.9844 | 0.7395 |
| 13 | 140.52 | 1.0072 | 0.9401 | 0.7078 |
| 14 | 140.44 | 1.0356 | 0.9656 | 0.7211 |
| 15 | 140.35 | 1.0620 | 0.9889 | 0.7331 |
| 16 | 140.35 | 1.063 | 0.9899 | 0.7335 |
| 17 | 140.35 | 1.0666 | 0.9930 | 0.7363 |
| 18 | 140.29 | 1.0425 | 0.9723 | 0.7074 |
| 19 | 140.29 | 1.0491 | 0.9749 | 0.7252 |
| 20 | 140.55 | 1.0357 | 0.9617 | 0.7198 |
| 21 | 140.49 | 0.9638 | 0.8912 | 0.6771 |
| 22 | 140.51 | 1.0491 | 0.9718 | 0.7069 |
| 23 | 140.51 | 1.0416 | 0.9715 | 0.7071 |
| 24 | 140.63 | 1.0071 | 0.9378 | 0.6867 |
| 25 | 140.87 | 0.9427 | 0.8765 | 0.6436 |
| 26 | 141.24 | 0.8426 | 0.7826 | 0.5816 |
| 27 | 140.55 | 1.0138 | 0.9425 | 0.6987 |
| 28 | 140.51 | 1.0418 | 0.9717 | 0.7059 |
| 29 | 140.51 | 1.0417 | 0.9718 | 0.7052 |
| 30 | 140.50 | 1.0447 | 0.9749 | 0.7065 |
| 31 | 140.50 | 1.0457 | 0.9759 | 0.7066 |
| 32 | 141.04 | 0.5567 | 0.5230 | 0.4068 |
| 33 | 141.38 | 0.3048 | 0.2876 | 0.2291 |
| 34 | 141.54 | 0.1893 | 0.1795 | 0.1462 |
| 35 | 141.42 | 0.2376 | 0.2276 | 0.1942 |
| 36 | 140.46 | 1.0757 | 1.0051 | 0.7245 |
| 37 | 140.43 | 1.1344 | 1.0647 | 0.7713 |
| 38 | 143.90 | 0.2568 | 0.2424 | 0.1831 |
| 39 | 141.86 | 0.8386 | **0.7894** | **0.5711** |
| 40 | 141.22 | 1.0064 | **0.9458** | **0.6423** |
| 41 | 140.87 | 1.0814 | 1.0138 | 0.6761 |
| 42 | 140.52 | 1.1447 | 1.0715 | 0.7238 |
| 43 | 140.51 | 1.1425 | 1.0694 | 0.7249 |
| 44 | 140.42 | 1.1119 | 1.0380 | 0.7260 |
| 45 | 140.43 | 1.0888 | 1.0168 | 0.7246 |
| 47 | 140.15 | 1.2449 | 1.1666 | 0.8064 |
| 48 | 140.15 | 0.7610 | 0.7143 | 0.5577 |
| 49 | 140.76 | 0.8818 | 0.8277 | 0.6455 |
| 50 | 140.59 | 0.9991 | 0.9356 | 0.7207 |
| 51 | 140.44 | 1.0002 | 0.9365 | 0.7208 |
| 52 | 140.47 | 1.0048 | 0.9396 | 0.7170 |
| 53 | 142.05 | 0.6603 | 0.6188 | 0.4649 |
| 54 | 140.59 | 1.0210 | 0.9526 | 0.6918 |
| 55 | 140.55 | 1.0301 | 0.9611 | 0.6976 |
| 56 | 140.55 | 1.0292 | 0.9602 | 0.6972 |
| 57 | 140.55 | 1.0284 | 0.9595 | 0.6968 |
| 58 | 140.61 | 0.9946 | 0.9275 | 0.6732 |
| 59 | 140.61 | 0.9906 | **0.9237** | **0.6711** |
| 60 | 140.61 | 0.9019 | **0.8475** | **0.6579** |
| 61 | 144.61 | 0 | 0 | 0 |
| 62 | 141.85 | 0 | 0 | 0 |
| 63 | 144.81 | 0 | 0 | 0 |
| 64 | 144.77 | 0 | 0 | 0 |
| 65 | 141.80 | 0 | 0 | 0 |
| 66 | 64.73 | 8.55 | 8.03 | 6.11 |
| 67 | 39.87 | 11.94 | 11.38 | 9.44 |
| 68 | 27.21 | 3.94 | 3.73 | 2.85 |
| 69 | 51.20 | 4.93 | 4.47 | 3.58 |
| 70 | 28.54 | 12.93 | 10.64 | 8.92 |

1-65: nodal heads [m] at nodes 1-65;

66-70: fixed-head nodes in/out flows [l/s] at nodes 61-65.

## VII. Conclusions

There were shown 2 CLA methods which involved a LS loop flows state estimator: ESM technique and a novel EM technique. Both techniques were applied consistently on two realistic water distribution systems.

The numerical results determined for the first water distribution system were verified with the numerical results obtained with an established CLA algorithm, namely the sensitivity matrix technique, which is based on the LS nodal heads state estimator.

The EM technique is by far more advantageous when judging in terms of computational time and also as opposed to the ESM technique. The confidence limits determined with the EM technique are the same as the confidence limits obtained with the ESM technique and the CLA technique (sensitivity matrix algorithm) based on the LS nodal heads state estimator.

It is demonstrated that the EM technique is able to employ efficiently the LS loop flows state estimator as there is not needed more than one simulation/state estimation for any water distribution system for the purposes of determining the confidence limits for the water distribution system state variables.

Because of the drastically reduced computational time, the novel EM technique is applicable to be used in on-line decision support systems of water networks [27] which can detect also malfunctions in water distribution systems based for example on pattern recognition and Artificial Neural Networks (ANNs). ANNs are mathematical models [113-115] more complex than simple regression models such as the response surface methods [116] and which ANNs show some degree of similarity with the biological correspondents.

The EM demonstrated also to be capable of making use of any associations of real pipe flows and pressure measurements with the scope of calculating the confidence limits during the tests that took place on a second medium size water distribution system.

APPENDIX

1.

The water network from Figure 1 is a realistic water network which belongs to a hilly area in UK and compares well in size with other water networks which can be found in hilly areas and which water networks are characterized by numerous pressure controlling valves, check valves and pumps. However, for simplicity in this paper the valves are replaced with pipes. Finally, the data for the water network consists of 34 nodes, 47 pipes and 5 inflow points and 3 outflow points. The inflow/outflow points are the reservoirs at nodes 27, 28, 29, 30, 31, 32, 33 and 34. Pipe data consisting of length [m], diameter [m] and C-values (conductivity) are listed in Table 1. This data, together with the reference pressure measurement in node 30 and five inflow and three outflow measurements allows calculation of the system state (a 42-dimensional vector of 34 nodal pressures and the 8 inflows/outflows in the fixed-pressure nodes). The head delivered by the pumps is called head pump and is function of the flow through the pump and can be described with the following equation:

$$h_p = A_P - B_p Q^{C_P} \qquad (22)$$

where $h_P$ is the pump head, $A_p$ is the pump head at zero flow, $Q$ is the flow through the pump and $B_P$, $C_P$ are the coefficients describing the pump curve shape.

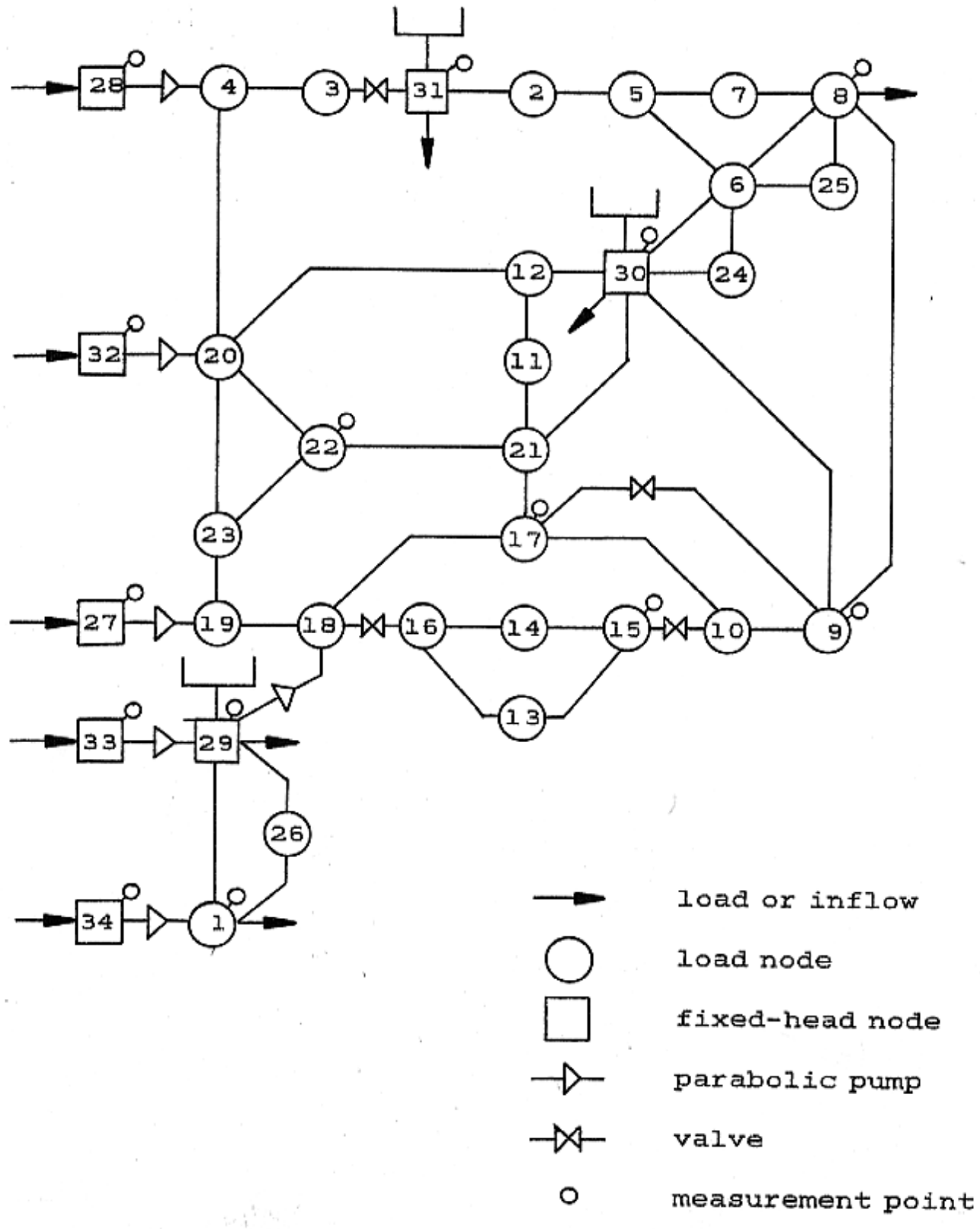

Figure 9. Water network found in hilly areas.

| Pipe | | Length | Diameter | C |
|---|---|---|---|---|
| Node | Node | | | |
| 3 | 4 | 606.6 | 0.4572 | 110 |
| 4 | 20 | 454.2 | 0.4572 | 110 |
| 20 | 23 | 2782.8 | 0.2286 | 105 |
| 19 | 23 | 304.8 | 0.3810 | 135 |
| 12 | 20 | 3383.3 | 0.3048 | 105 |
| 20 | 22 | 1767.8 | 0.4572 | 110 |
| 22 | 23 | 1015.0 | 0.3810 | 135 |
| 18 | 19 | 1097.3 | 0.3810 | 135 |
| 2 | 31 | 3150.6 | 0.3048 | 100 |
| 21 | 22 | 762.0 | 0.4572 | 110 |
| 17 | 18 | 914.4 | 0.2286 | 125 |
| 14 | 16 | 411.5 | 0.1524 | 100 |
| 14 | 15 | 701.0 | 0.2286 | 110 |
| 13 | 16 | 1072.9 | 0.2286 | 135 |
| 13 | 15 | 864.1 | 0.1524 | 90 |
| 10 | 17 | 832.1 | 0.1524 | 90 |
| 17 | 21 | 1969.4 | 0.2286 | 95 |
| 11 | 21 | 777.2 | 0.2286 | 90 |
| 11 | 12 | 542.5 | 0.2286 | 90 |
| 21 | 30 | 1600.2 | 0.4572 | 110 |
| 12 | 30 | 249.9 | 0.3048 | 105 |
| 2 | 5 | 1028.7 | 0.2286 | 110 |
| 24 | 30 | 443.7 | 0.2286 | 90 |
| 6 | 30 | 743.7 | 0.3810 | 100 |
| 9 | 30 | 931.1 | 0.2286 | 125 |
| 9 | 10 | 2689.9 | 0.1524 | 100 |
| 5 | 7 | 326.1 | 0.1524 | 100 |
| 7 | 8 | 844.3 | 0.2286 | 110 |
| 5 | 6 | 1274.0 | 0.1524 | 100 |
| 6 | 8 | 1115.6 | 0.2286 | 90 |
| 6 | 25 | 615.5 | 0.3810 | 110 |
| 8 | 9 | 1406.7 | 0.1524 | 100 |
| 1 | 29 | 426.7 | 0.2540 | 100 |
| 1 | 26 | 2098.1 | 0.3556 | 100 |
| 8 | 25 | 500.0 | 0.3810 | 110 |
| 6 | 24 | 300.0 | 0.2286 | 90 |
| 26 | 29 | 1500 | 0.3556 | 100 |
| 3 | 31 | 1930.9 | 0.4572 | 110 |
| 9 | 17 | 2334.8 | 0.1524 | 100 |
| 16 | 18 | 823.0 | 0.3048 | 140 |
| 10 | 15 | 711.7 | 0.1524 | 90 |

| Pumps | | Ap | Bp | Cp |
|---|---|---|---|---|
| 28 | 4 | 122.44 | 2.01 | 0.004792 |
| 32 | 20 | 102.42 | 1.04 | 0.3296 |
| 27 | 19 | 62.03 | 1.67 | 0.002128 |
| 29 | 18 | 18.89 | 2.0 | 0.004001 |
| 34 | 1 | 44.5 | 1.64 | 0.002406 |
| 33 | 29 | 75.47 | 2.38 | 0.000486 |

Table 8. Pipe data for the water network.

| **Node consumptions** | **[l/s]** | **Node consumptions** | **[l/s]** | **Reference pressure [m]** | |
|---|---|---|---|---|---|
| 1 | 52.6 | 18 | 12.1 | 30 | 43.5619 |
| 2 | 2.7 | 19 | 4.5 | **Inflows [l/s]** | |
| 3 | 19.2 | 20 | 12.1 | Node | [l/s] |
| 4 | 5.9 | 21 | 22.3 | 27 | 34.0 |
| 5 | 1.1 | 22 | 32.4 | 28 | 96.5 |
| 6 | 2.1 | 23 | 38.2 | 29 | 64.3 |
| 7 | 3.0 | 24 | 5.0 | 30 | 106.3 |
| 8 | 69.4 | 25 | 9.0 | 31 | 38.9 |
| 9 | 8.1 | 26 | 11.1 | 32 | 6.0 |
| 10 | 3.8 | 27 | 6.2 | 33 | 121.7 |
| 11 | 1.9 | 28 | 0 | 34 | 21.6 |
| 12 | 10.2 | 29 | 22.9 | | |
| 13 | 21.2 | 30 | 39.5 | | |
| 14 | 10.3 | 31 | 39.3 | | |
| 15 | 22.2 | 32 | 0 | | |
| 16 | 4.7 | 33 | 0 | | |
| 17 | 2.4 | 34 | 0 | | |

Table 9. Measurement data.

2.

The water network from Figure 1 is a realistic water network which belongs to a city in UK and compares well in size with other water networks used presently in modeling and simulation. The data for the water network consists of 65 nodes, 91 pipes and 5 inflow points. The inflow points are the reservoirs at nodes 61, 62, 63, 64 and 65. Pipe data consisting of length [m], diameter [m] and C-values (conductivity) are listed in Table 1. This data, together with the reference pressure measurement in node 64 and five inflow measurements and fixed head values in nodes 61, 62, 63, 64, and 65 allows calculation of the system state (a 70-dimensional vector of 65 nodal pressures and the 5 inflows in the fixed-pressure nodes). Measurement data is shown in Table 2.

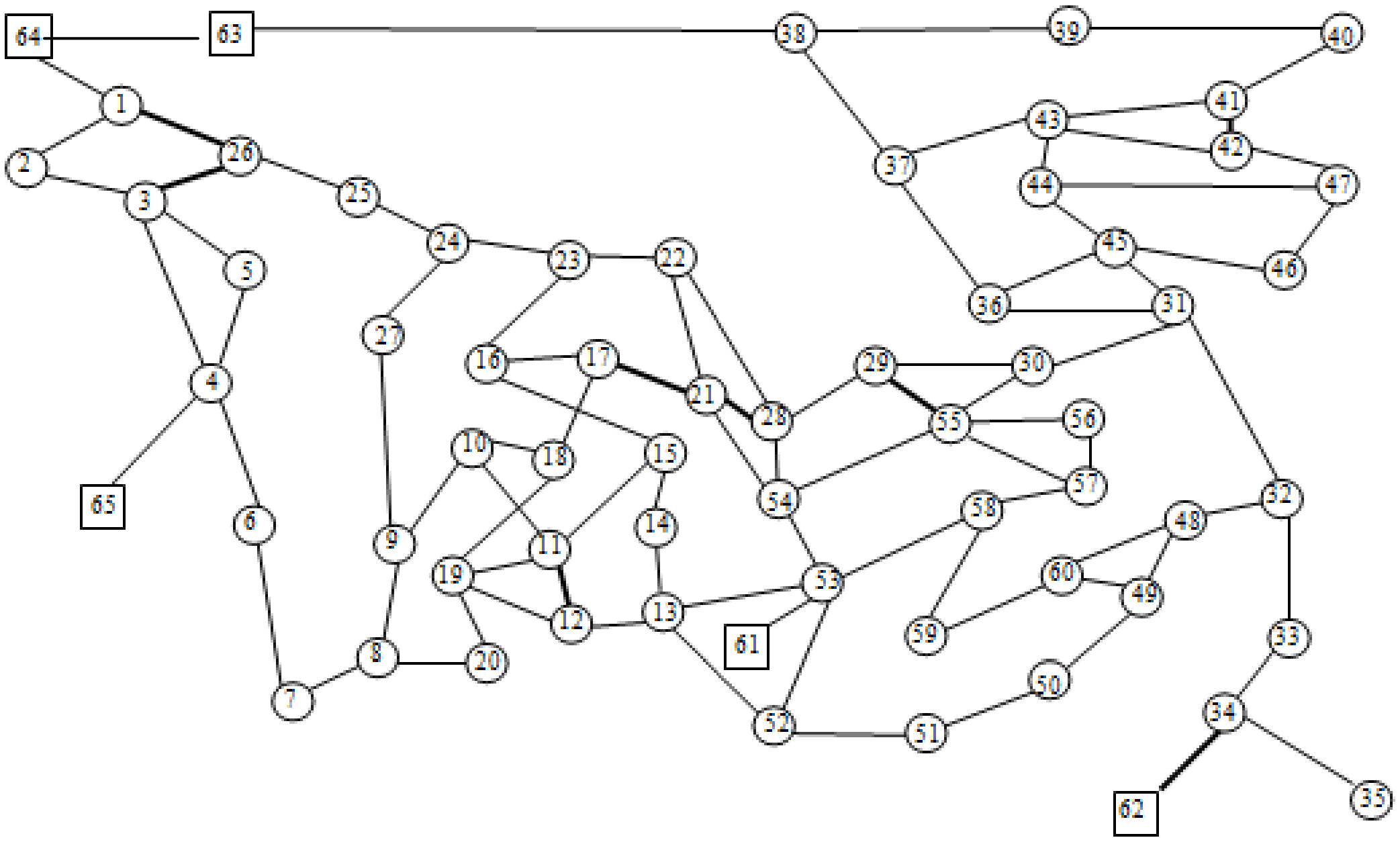

Figure 10. Water network

| Pipe | | Length | Diameter | C |
|---|---|---|---|---|
| Node | Node | | | |
| 64 | 1 | 370 | 0.381 | 50 |
| 1 | 2 | 350 | 0.225 | 110 |
| 2 | 3 | 770 | 0.225 | 110 |
| 3 | 4 | 800 | 0.225 | 110 |
| 4 | 5 | 210 | 0.125 | 60 |
| 4 | 6 | 270 | 0.225 | 160 |
| 7 | 6 | 220 | 0.225 | 160 |
| 8 | 7 | 480 | 0.3 | 158 |
| 9 | 8 | 380 | 0.225 | 159 |
| 10 | 9 | 190 | 0.225 | 145 |
| 11 | 10 | 550 | 0.225 | 145 |
| 12 | 11 | 610 | 0.225 | 145 |
| 13 | 12 | 780 | 0.225 | 80 |
| 14 | 13 | 320 | 0.225 | 119 |
| 15 | 14 | 710 | 0.25 | 90 |
| 16 | 15 | 230 | 0.25 | 80 |
| 17 | 16 | 380 | 0.25 | 80 |
| 18 | 17 | 320 | 0.168 | 120 |
| 19 | 18 | 580 | 0.168 | 120 |
| 20 | 19 | 1060 | 0.175 | 80 |
| 21 | 17 | 310 | 0.25 | 70 |
| 22 | 21 | 270 | 0.2 | 145 |
| 23 | 22 | 430 | 0.3 | 145 |
| 24 | 23 | 250 | 0.3 | 85 |
| 25 | 24 | 260 | 0.3 | 85 |
| 26 | 25 | 730 | 0.3 | 118 |
| 27 | 24 | 720 | 0.2 | 145 |
| 28 | 22 | 550 | 0.3 | 229 |
| 29 | 28 | 210 | 0.3 | 127 |
| 30 | 29 | 147 | 0.3 | 145 |
| 31 | 30 | 120 | 0.3 | 145 |
| 32 | 31 | 1510 | 0.225 | 96 |
| 33 | 32 | 970 | 0.3 | 165 |
| 34 | 33 | 400 | 0.3 | 165 |
| 35 | 34 | 800 | 0.2 | 140 |
| 36 | 31 | 160 | 0.225 | 80 |
| 37 | 36 | 600 | 0.225 | 60 |
| 38 | 37 | 1770 | 0.225 | 47 |
| 39 | 38 | 3090 | 0.356 | 46 |
| 40 | 39 | 410 | 0.225 | 80 |
| 41 | 40 | 420 | 0.225 | 80 |
| 42 | 41 | 400 | 0.15 | 145 |
| 43 | 42 | 150 | 0.15 | 116 |
| 44 | 43 | 460 | 0.094 | 170 |
| 45 | 44 | 530 | 0.15 | 145 |
| 46 | 45 | 1220 | 0.15 | 139 |
| 47 | 46 | 600 | 0.15 | 145 |
| 48 | 32 | 300 | 0.225 | 135 |
| 49 | 48 | 350 | 0.225 | 171 |
| 50 | 49 | 710 | 0.225 | 110 |
| 51 | 50 | 225 | 0.225 | 110 |
| 52 | 51 | 310 | 0.225 | 90 |
| 53 | 52 | 590 | 0.094 | 80 |
| 54 | 53 | 650 | 0.2 | 158 |
| 55 | 54 | 300 | 0.3 | 145 |
| 56 | 55 | 430 | 0.3 | 145 |
| 57 | 56 | 330 | 0.3 | 145 |
| 58 | 57 | 250 | 0.225 | 158 |
| 59 | 58 | 200 | 0.175 | 115 |
| 60 | 59 | 180 | 0.125 | 60 |
| 61 | 53 | 360 | 0.3 | 80 |
| 62 | 34 | 400 | 0.3 | 165 |
| 38 | 63 | 2200 | 0.356 | 100 |
| 48 | 60 | 350 | 0.125 | 60 |
| 4 | 65 | 440 | 0.3 | 170 |
| 60 | 49 | 360 | 0.15 | 105 |
| 53 | 58 | 740 | 0.175 | 115 |
| 30 | 55 | 420 | 0.15 | 90 |
| 53 | 13 | 330 | 0.175 | 127 |
| 13 | 52 | 630 | 0.225 | 104 |
| 15 | 11 | 520 | 0.25 | 70 |
| 11 | 19 | 540 | 0.25 | 70 |
| 19 | 12 | 900 | 0.175 | 80 |
| 20 | 8 | 660 | 0.225 | 140 |
| 18 | 10 | 610 | 0.125 | 55 |
| 21 | 54 | 390 | 0.2 | 145 |
| 21 | 28 | 280 | 0.168 | 120 |
| 28 | 54 | 300 | 0.15 | 90 |
| 29 | 55 | 180 | 0.3 | 112 |
| 36 | 45 | 340 | 0.15 | 90 |
| 45 | 31 | 390 | 0.2 | 145 |
| 47 | 42 | 670 | 0.15 | 116 |
| 43 | 37 | 1100 | 0.142 | 105 |
| 41 | 43 | 350 | 0.094 | 170 |
| 44 | 47 | 1470 | 0.15 | 81 |
| 1 | 26 | 630 | 0.3 | 50 |
| 23 | 16 | 720 | 0.25 | 80 |
| 27 | 9 | 700 | 0.142 | 137 |
| 5 | 3 | 590 | 0.125 | 60 |
| 3 | 26 | 2050 | 0.15 | 40 |
| 55 | 57 | 760 | 0.117 | 60 |
| 63 | 64 | 270 | 0.381 | 32 |

Table 10. Pipe data for the water network.

| **Node consumptions** | **[l/s]** | **Node consumptions** | **[l/s]** | **Node consumptions** | **[l/s]** |
|---|---|---|---|---|---|
| **1** | 2.88 | **32** | 1.64 | **56** | 0.57 |
| **2** | 2.94 | **33** | 2.09 | **57** | 8.71 |
| **3** | 10.46 | **34** | 6.77 | **58** | 0.5 |
| **4** | 2.15 | **35** | 4.85 | **59** | 0.35 |
| **5** | 3.84 | **36** | 1.91 | **60** | 0.34 |
| **6** | 1.87 | **37** | 11.51 | **61** | 0.26 |
| **7** | 0.67 | **38** | 2.81 | **62** | 7.95 |
| **8** | 4.54 | **39** | 2.77 | **63** | 0.58 |
| **9** | 10.83 | **40** | 3.75 | **64** | 0 |
| **10** | 0.78 | **41** | 1.32 | **65** | 2.46 |
| **11** | 10.80 | **42** | 1.16 | **Inflows [l/s]** | |
| **12** | 7.40 | **43** | 1.35 | **61** | 64.73 |
| **13** | 2.71 | **44** | 0.42 | **62** | 39.87 |
| **14** | 2.58 | **45** | 2.64 | **63** | 27.21 |
| **15** | 1.89 | **46** | 2.79 | **64** | 51.20 |
| **16** | 3.15 | **47** | 5.37 | **65** | 28.54 |
| **17** | 6.77 | **48** | 1.16 | **Reference pressure** | |
| **18** | 2.98 | **49** | 9.64 | **64** | 144.77 |
| **19** | 1.1 | **50** | 6.54 | **Fixed nodal head** | |
| **20** | 2.95 | **51** | 3.18 | **61** | 144.61 |
| **21** | 2.13 | **52** | 2.01 | **62** | 141.85 |
| **22** | 2.36 | **53** | 8.51 | **63** | 144.81 |
| **23** | 8.03 | **54** | 1.73 | **64** | 144.77 |
| **24** | 2.73 | **55** | 0.47 | **65** | 141.80 |
| **25** | 0.16 | **56** | 0 | | |
| **26** | 5.68 | **57** | 8.71 | | |
| **27** | 1.74 | **58** | 0.5 | | |
| **28** | 0.65 | **59** | 0.35 | | |
| **29** | 1.92 | **60** | 0.34 | | |
| **30** | 4.52 | **61** | 0.26 | | |
| **31** | 2.35 | **62** | 7.95 | | |

Table 11. Measurement data.